\documentclass[11pt,english]{article}

\usepackage[T1]{fontenc}
\usepackage[utf8]{inputenc}
\usepackage{lmodern}
\usepackage{amsmath}
\usepackage{amssymb}
\usepackage{mathrsfs}
\usepackage{cancel}
\usepackage[square,numbers,sort&compress]{natbib}
\usepackage[english]{babel}
\usepackage{physics}

\usepackage{geometry}
\usepackage{caption}
\usepackage{subcaption}
\usepackage{graphicx}
\usepackage{physics}
\graphicspath{{Figures/}}
\usepackage[toc,page]{appendix}

\usepackage{caption}
\usepackage{epstopdf} 
\usepackage{comment}
\usepackage{tabularx} 
\usepackage{booktabs} 
\usepackage{psfrag} 

\usepackage{eurosym} 

\usepackage{xcolor} 
\usepackage{empheq} 

\definecolor{linkcolor}{rgb}{0,0,0.6} 
\usepackage[ 
    pdftex,
    colorlinks=true,
    pdfstartview=FitV,
 linkcolor=linkcolor,
    citecolor=linkcolor,
    urlcolor=linkcolor,
    hyperindex=true,
    hyperfigures=false,
    backref=page,            
]{hyperref}
\renewcommand*{\backref}[1]{}
\renewcommand*{\backrefalt}[4]{%
  \ifcase #1 
  \or
    (cit. on p.~#2)%
  \else
    (cit. on pp.~#2)%
  \fi
}

\usepackage{fancyhdr} 
\usepackage{pgfplots} 
\usepgfplotslibrary{groupplots}

\newcommand\mv{\sigma}
\newcommand\Ne{\mathscr{E}_1}
\newcommand\alg{\mathscr{E}_2}

\newcommand\rmax{r_*^{\max}}
\newcommand\phiback{\bar{\phi}}
\newcommand\gmunuback{\bar{g}_{\mu \nu}}

\newcommand\Veff{V_{\mathrm{eff}}}

\newcommand\mreg{M_p^{\mathrm{reg}}} 
\newcommand\Bterm{\mathscr{B}}
\newcommand\Dope{\mathscr{D}}
\newcommand\Schroop{{\mathscr{L}}}
\newcommand\SchroopE{\Schroop_{E}}

\newcommand\cA{\mathcal{A}}
\newcommand\cB{\mathcal{B}}
\newcommand\cC{{\mathcal{C}}}

\newcommand\nth{{p}}
\newcommand\Ftwo{{F}}

\newcommand\cCo{{C}}
\newcommand\cDis{D}
\newcommand\XH{\tilde X}

\newcommand\Det{\Delta}
\newcommand\Znew{\cal Z}
\newcommand\Ynew{\cal Y}
\newcommand\reff{r_0}
\newcommand\rsc{r_m}

\numberwithin{equation}{section}
\numberwithin{figure}{section}
\numberwithin{table}{section}

\newcolumntype{Y}{>{\centering\arraybackslash}X}
\pgfplotsset{compat=1.6}
\pgfplotsset{/pgf/number format/use comma}

\author{Christos Charmousis$^{a}$, Simon Iteanu$^{a}$, David Langlois$^{b}$ and Karim Noui$^{a}$ \\
\small $^{a}${\it{Universit\'e Paris-Saclay, CNRS/IN2P3, IJCLab, 91405 Orsay, France}}\\
\small $^{b}${\it{Université Paris Cit\'e, CNRS, Astroparticule et Cosmologie, F-75013 Paris, France}}
}

\date{}
\title{Radial  Perturbations of  Black Holes in DHOST Theories}

\begin{document}

\maketitle
\begin{center}
{\bf Abstract}
\end{center}
We study radial perturbations of static black holes with primary hair in a subfamily of degenerate higher-order scalar-tensor (DHOST) theories. We recast the equation of motion for the monopole degree of freedom into a flat radial wave equation and show that the associated operator can be extended, through appropriate boundary conditions, to a positive self-adjoint operator which ensures the stability of the radial mode. Remarkably, the coordinate choice that leads to the flat wave equation corresponds to the unitary gauge, in which the scalar field is uniform. As a result, the radial coordinate extends beyond the event horizon, into the black hole interior, in contrast with the tortoise coordinate in General Relativity.  The same wave equation with the same coordinate choice  applies to all solutions that are connected by disformal transformations. 
We also examine stealth black hole solutions, with either a constant or non constant kinetic term. In the former case, we find, to linear order, the absence of a propagating degree of freedom. In the latter case, we identify a stable radial degree of freedom, except for special values of the theory coupling constants.

\newpage
\tableofcontents
\newpage
\section{Introduction}

Future observational data on black holes (BH) will offer an unprecedented window into the behavior of gravity in the strong-field regime, particularly in the vicinity of compact objects. Black holes, in this context, stand out as ideal laboratories for testing General Relativity (GR) and exploring potential deviations from its predictions.

Several complementary approaches exist to test GR. The most direct method involves verifying the internal consistency of observational data against GR’s predictions. A more challenging alternative involves constructing and analyzing alternative theories, then comparing their observational signatures to those of GR. Due to the complexity of this task, only a limited number of alternative theories have been investigated so far. Nevertheless, even qualitative differences arising from these theories can inspire novel methods for analyzing future data. Many modified gravity models introduce a single scalar degree of freedom in addition to the standard tensor modes. The most general theories of this type, known as Degenerate Higher-Order Scalar-Tensor (DHOST) theories~\cite{Langlois:2015cwa,Langlois:2015skt,BenAchour:2016fzp} (see e.g. \cite{Langlois:2018dxi,Kobayashi:2019hrl} for reviews), encompass the scalar-tensor theories typically studied in the literature.

As the development of GR has shown, the existence of exact solutions often provides crucial insights into the subtle aspects of gravitational behavior. This underscores the importance of identifying exact BH solutions  beyond GR. Finding explicit solutions in modified gravity, particularly within DHOST theories, is a technically demanding process (for a recent review see for example \cite{Babichev:2023psy, Lecoeur:2024kwe}). The earliest and simplest such solutions were stealth solutions \cite{Babichev:2013cya, Kobayashi:2014eva,Minamitsuji:2019shy} where the metric remains identical to that of GR, but the scalar field exhibits a non-trivial profile. Initially static, these solutions were later extended to include the first rotating solution, stealth Kerr~\cite{Charmousis:2019vnf}, which was subsequently shown to correspond to a genuinely distinct black hole geometry, the disformed Kerr metric~\cite{Anson:2020trg,BenAchour:2020fgy} (see also \cite{BenAchour:2025lkx}). Perturbations around these solutions obtained in various theories have been considerably studied in recent years,  using different approaches (see e.g. \cite{Tomikawa:2021pca, deRham:2019gha,Khoury:2020aya,Langlois:2021xzq,Kobayashi:2012kh,Cisterna:2015uya,Takahashi:2016dnv,Takahashi:2019oxz,Charmousis:2019fre,Langlois:2021aji,Babichev:2017lmw,Babichev:2018uiw,Takahashi:2021bml,Chatzifotis:2021pak,Minamitsuji:2022vbi,Langlois:2022eta,Minamitsuji:2022mlv,Noui:2023ksf,Roussille:2023sdr,Antoniou:2024gdf,Antoniou:2024hlf,Franciolini:2018uyq,Hui:2021cpm,Mukohyama:2022enj,Mukohyama:2023xyf,Mukohyama:2025owu,Sirera:2024ghv,Smulders:2026qwc,Konoplya:2026cus}).

Recently, well-defined static black holes with {\it primary hair} have been obtained within a subfamily of shift-symmetric DHOST theories~\cite{Bakopoulos:2023fmv} (see also the extensions found in \cite{ Baake:2023zsq, Bakopoulos:2023sdm}). By primary hair, we mean solutions where the metric depends on an independent parameter associated with the scalar field, in addition to the usual ADM mass. When this scalar parameter is set to zero, the Schwarzschild metric is recovered, indicating that these solutions represent one-parameter deformations of the standard GR metric. The key property allowing a scalar charge is the presence of global shift symmetry of the theory which in turn allows for a linearly time dependent scalar field firstly introduced in \cite{Babichev:2013cya}. This mild time dependence yields in turn a regular scalar solution for arbitrary values of the scalar charge \cite{Bakopoulos:2023fmv} allowing the hair to be primary rather than secondary.

A critical question concerning these solutions is their stability. In prior work, we examined their axial perturbations \cite{Charmousis:2025xug}, which are simpler than polar perturbations because they involve only a single degree of freedom, as in GR. Notably, we exploited the property that the dynamics of axial perturbations can be formally related to the GR equations of motion for axial perturbations in an effective metric as shown in \cite{Langlois:2022ulw} (see also \cite{Tomikawa:2021pca}). The causal structure of this effective metric differs from that of the background metric, leading to distinct gravitational and luminous horizons. Using a WKB approximation, we computed the quasi-normal modes for the Schrödinger-like equation associated with the effective metric outside the gravitational horizon.

In this work, we focus on the {\it radial} perturbations of BHs  with primary hair, as well as a selection of other black hole solutions discussed in the literature \cite{Charmousis:2021npl,Minamitsuji:2019shy}. Like axial modes, radial perturbations also depend on a single degree of freedom—in contrast with GR where radial perturbations are absent  as a consequence of Birkhoff’s theorem. As we demonstrate, the dynamics of this degree of freedom, which combines both scalar and metric perturbations, can be described by a simple Schrödinger-like equation. To achieve this, we introduce new time and radial coordinates that simplify the equation of motion into a Schrödinger-like form for a master variable.

Applying this framework to black hole solutions with primary hair, we find a remarkable result: the new time coordinate coincides with the scalar field itself, while the new radial coordinate is confined to a finite interval (unlike the tortoise coordinate in GR, which spans the entire real line). We then derive the corresponding potential, which is not always positive, and discuss the stability of the perturbations—a question tied to the existence of a positive self-adjoint operator.

From any black hole solution in a given DHOST theory, new solutions can be generated via conformal-disformal transformations of the metric~\cite{BenAchour:2016cay}. These transformations are field redefinitions of the metric that depend on the scalar field. The disformed metric represents a genuinely new black hole solution associated with a different DHOST theory, assuming ordinary matter and light are minimally coupled to the disformed metric. As an illustration, we consider black hole solutions with primary hair in Horndeski theory, obtained through disformal transformation of the original solutions. We find that the Schrödinger-like equation governing the radial perturbations of the disformed solution is identical to that of the original solution, with the same new time and radial coordinates. This invariance under disformal transformations was also observed for axial perturbations.

We also examine radial perturbations for  stealth solutions, i.e. Schwarzschild black holes with a non trivial scalar field profile, within the context of Horndeski theories. For  stealth solutions with a non-constant kinetic term $X\equiv -\partial_\mu\phi \,\partial^\mu\phi/2$, obtained using the same approach as for BHs with primary hair, we find that radial perturbations 
do not yield 
a propagating degree of freedom at the linear level.
 We also revisit stealth Schwarzschild black hole perturbations characterized by a constant $X$, which have also been studied elsewhere \cite{Babichev:2018uiw, deRham:2019gha,Khoury:2020aya,Langlois:2021aji}.

The paper is organized as follows. In the next section, we introduce our general framework and present our systematic procedure for extracting the degree of freedom associated with radial oscillations and formulating its equation of motion in a Schr\"odinger-like form.   Section \ref{sec.Primary_hair_BH} recaps the BH solutions with primary hair and applies our procedure to study their radial oscillations. In section \ref{section_Horndeski}, we focus on BHs within Horndeski theories, in particular stealth solutions with both constant $X$ and non constant $X$.
The final section summarises our results. Additional technical details are provided in the appendices.

\section{Theoretical framework and radial perturbations}
In this Section, we present the general procedure to compute, from the perturbation equations about an arbitrary background in DHOST theories, a Schr\"odinger-like equation for the radial perturbations. 

\subsection{DHOST theories and background solutions}
We consider  DHOST theories (up to quadratic terms in the second derivatives of the scalar field) that are described by the action \cite{Langlois:2015cwa}
\begin{align}
\label{DHOSTaction}
 S\left[g_{\mu \nu}, \phi\right]= \int \mathrm{d}^4 x \sqrt{-g} \left( P(X, \phi)+ Q(X, \phi) \square \phi+\Ftwo(X, \phi) R +  \sum_{i=1}^5 A_i(X, \phi) L_i^{(2)} \right)  , 
\end{align}
where the kinetic density is defined by
\begin{equation}
    X\equiv  -\frac{1}{2} \partial_\mu \phi \, \partial^\mu \phi\,.
\end{equation}
and the five elementary quadratic Lagrangians are given by
\cite{Langlois:2015cwa} 
\begin{eqnarray*}
&L_{1}^{(2)}=
\phi^{\mu\nu}\phi_{\mu\nu}\,, \qquad L_{2}^{(2)}=\left( \square\phi\right)^{2}\,, \qquad L_{3}^{(2)}= (\square\phi)\phi^{\mu}\phi_{\mu\nu}\phi^\nu\,,
\qquad \\
  & L_{4}^{(2)}=\phi^\lambda\phi_{\lambda\mu}\phi^{\mu\nu}\phi_\nu \,, \qquad \qquad
  L_{5}^{(2)}=\left(\phi^{\mu}\phi_{\mu\nu}\phi^\nu\right)^{2}\,,
  \qquad
\end{eqnarray*}
using the  notation $\phi_\mu\equiv \nabla_{\mu}\phi$ and $\phi_{\mu\nu}\equiv \nabla_{\nu}\nabla_{\mu}\phi$.
 In the above action, $P$, $Q$, $\Ftwo$ and $A_i$ are functions of $\phi$ and $X$ and must satisfy three so-called degeneracy conditions so that the theory contains a single scalar degree of freedom. 

The theories that will be discussed in this paper all belong to a sub-class of
DHOST theories, known as Beyond Horndeski theories~\cite{Gleyzes:2014dya,Gleyzes:2014qga}, characterised by the conditions
\begin{eqnarray}
\label{Ai_GLPV}
A_2=-A_1   \,, \qquad A_4=-A_3= \frac{A_1+\Ftwo_{X}}{X} \,, \qquad A_5=0\,.
\end{eqnarray}
This sub-class contains the Horndeski theories~\cite{Horndeski:1974wa}, which satisfy the more restrictive conditions
\begin{eqnarray}
\label{Ai_Horndeski}
A_2=-A_1= \Ftwo_{X}  \,, \qquad A_3=A_4=A_5=0\,.
\end{eqnarray}
From now on, we restrict our attention to Lagrangians in which $Q=0$ and the remaining independent functions $P$, $\Ftwo$ and $A_1$ depend only on $X$, which implies in particular that  the theory has shift ($\phi \rightarrow \phi + c$ where $c$ is a constant) and parity global symmetries ($\phi \rightarrow -\phi$) for the scalar field $\phi$.

The background which we shall be perturbing are static and spherically symmetric solutions to the equations of motion, 
\begin{align}
\label{line_element}
\mathrm{d} s^2 &=-\cA(r) \mathrm{d} t^2+\frac{\mathrm{d} r^2}{\cB(r)} + r^2 \mathrm{d} \Omega^2, \qquad \mathrm{~d} \Omega^2=\mathrm{d} \theta^2+\sin ^2 \theta \mathrm{d} \varphi^2 \, ,
\end{align}
with a  scalar field admitting a linear time dependence (allowed by the shift symmetry),
\begin{equation}
\label{phiback}
    \phiback=q t+\psi(r) \, .
\end{equation}

\subsection{Perturbation theory and gauge fixing}
We now consider linear perturbations about the background solution \eqref{line_element}-\eqref{phiback}. Let $h_{\mu \nu}$ be the perturbation about the background metric $\gmunuback$ and $\delta \phi$ the scalar perturbations about the background scalar field  $\phiback$:
\begin{equation}
   h_{\mu \nu} =g_{\mu \nu}-\gmunuback, \qquad \delta \phi=\phi-\phiback.
   \label{pertgeneral}
\end{equation}
 Due to the spherical symmetry of the background, we express both perturbations as expansions in spherical harmonics $Y_{\ell m}(\theta, \varphi)$, and separate them into axial (odd-parity) and polar (even-parity) modes as the two sectors are decoupled at linear order.
We have already  studied axial modes in a previous work \cite{Langlois:2022ulw}.
For a general $\ell\geq2$, the polar modes 
are described  by seven components of the metric perturbation and the perturbation of the scalar field. In this article we only study  the monopole $\ell=0$  where the  non-trivial components of the perturbations \eqref{pertgeneral} can be parametrized as follows,
\begin{align}
&h_{t t}  =\cA(r)  H_0(t, r) \, ,\quad
h_{t r} = H_1(t, r) \, , \quad 
h_{r r}  =\frac{1}{\cA(r)}  H_2(t, r) \, , \\
&h_{a b}  = K(t, r)\,  \bar{g}_{a b}
\quad (a,b \in \{\theta,\varphi \})
\,, \qquad
\delta \phi= \delta \phi(t, r) \, .
\end{align}
In the following, we adopt the gauge conditions 
 \begin{equation}
 \label{gauge}
     K=0,\qquad \delta \phi=0 \, ,
 \end{equation}
 which completely fix the gauge,
  and derive the set of equations of motion for the three remaining variables: $H_0$,  $H_1$ and $H_2$.
As shown in \eqref{Gauge_Fixing_T}, one can fix $\delta\phi=0$ only when $q$ is non zero.
\subsection{Master equation for the monopole perturbation}
\label{perturbations section}
Let us now outline the general strategy for the construction of the master variable describing the monopole perturbation and the corresponding second-order equation. 
First, we observe that, among the ten linearized metric equations $\mathscr{E}_{\mu\nu}=0$, many are trivially satisfied, and we can extract three independent equations corresponding to the components
$\mathscr{E}_{tt}$, $\mathscr{E}_{tr}$ and $\mathscr{E}_{rr}$\footnote{As shown in App.\ref{App:Perturbation_Equation}, the other non-vanishing equations are linear combinations of the three aforementioned equations and their derivatives.}. These three equations are linear combinations of the $H_n$ and their derivatives $\dot H_n$, $H_n'$ (with $n=0,1,2$) and take the schematic form:
\begin{eqnarray}
\mathscr{E}_{tt} & = & [H_0,H_1,H_2,H_1',H_2',\dot H_2] \, ,\label{tt_equation} \\
\mathscr{E}_{tr} & = & [H_0,H_1,H_2,H_0',\dot H_2] \, , \label{tr_equation} \\
\mathscr{E}_{rr} & = & [H_0,H_1,H_2,H_0',\dot H_0,\dot H_1] \, , \label{rr_equation}
\end{eqnarray}
Here and throughout a prime denotes the derivative with respect to $r$ while a dot the derivative with respect to $t$, and the notation $[\cdots]$ means a linear combination of the terms enclosed  within the brackets with coefficients that depend on the background solution. Their general explicit expressions are not needed here.

Next, we combine \eqref{tt_equation} and \eqref{tr_equation} 
to eliminate $\dot H_2$, which gives a new equation of the form
\begin{eqnarray}
    \Ne = \;
\tilde a^{(0)}\, H_0
+\tilde a^{(1)}\, H_1
+\tilde a^{(2)}\, H_2
+\tilde b^{(0)}\, H_0'
+\tilde b^{(1)}\, H_1'
+\tilde b^{(2)}\, H_2' \, ,
\end{eqnarray}
We then introduce the new function
\begin{equation}
\mv= \tilde b^{(0)}\, H_0
+\tilde b^{(1)}\, H_1
+\tilde b^{(2)}\, H_2 \,.
\label{nnvv}
\end{equation}
so that the previous equation can be rewritten in the form
   \begin{eqnarray}
    \Ne = \; \alpha_1^{(0)}\, H_0
+\alpha_1^{(1)}\, H_1
+\beta_1\, \sigma
+\gamma_1\, \sigma' \, ,
\label{EqE1}
\end{eqnarray} 
where $H_2$ has been expressed in terms of $\sigma$, $H_0$ and $H_1$, using \eqref{nnvv}.

Similarly, combining $\mathscr{E}_{tr}$ and $\mathscr{E}_{rr}$ to eliminate $H_0'$, we obtain a new equation which, remarkably\footnote{These properties are a  consequence of some general relations (for a solution of the background equations) between the coefficients entering in the linear combinations \eqref{tt_equation} and \eqref{tr_equation}.
},
can be written in the form
 \begin{eqnarray}
    \alg = \; \alpha_2^{(0)}\, H_0
+\alpha_2^{(1)}\, H_1
+\beta_2\, \sigma
+\gamma_2\, \dot \sigma \, . \label{EqE2}
\end{eqnarray} 
The system of equations $\{\Ne=0 \, ,\alg=0\}$ enables us  to express $H_0$ and $H_1$ in terms of
$\mv$ and its derivatives, provided  the associated determinant, 
\begin{equation}
\label{determinant_def}
\Det = \alpha_1^{(0)} \alpha_2^{(1)}  - \alpha_1^{(1)} \alpha_2^{(0)} \,,
\end{equation}
does not vanish. 
Notice that an example where the determinant $\Det$ vanishes will be treated as a special case later on.
Finally, substituting these expressions of $H_0$ and $H_1$ in terms of $\sigma$ and its derivatives, and  that of $H_2$ using \eqref{nnvv},  into one of the three original equations \eqref{tt_equation}, \eqref{tr_equation} or \eqref{rr_equation}, we obtain a second-order differential equation for the master variable $\mv(t,r)$ of the form
\begin{equation}
\label{def_second_order_equation}
    d_1 \, \mv''+d_2\, \ddot{\mv}+d_3 \, \dot{\mv'}+d_4 \, \mv'+d_5 \, \dot{\mv}+d_6 \, \mv=0 \, .
\end{equation}
The coefficients $d_i$ can be computed in terms of the original coefficients entering the metric equations \eqref{tt_equation}, \eqref{tr_equation} and \eqref{rr_equation} but their expressions are too cumbersome to be written here in full generality. We will give their explicit expressions in the following sections for some cases of interest.

\subsection{Schrödinger-like equation}
It is convenient to rewrite the master equation (\ref{def_second_order_equation}) in the form of  a Schrödinger-like equation, by introducing  new time and radial variables. 

First, one can get rid of both  $\dot{\mv}'$ and $\dot{\mv}$ in  \eqref{def_second_order_equation} 
by defining the new time coordinate
\begin{equation}
\label{def_tstar}
    t_*=t +\int W(r)\mathrm{d}r \,, \qquad W=-\frac{d_3}{2d_1}\,.
\end{equation}
Indeed, due to non-trivial relations between the equation coefficients $d_i$, $W$ also satisfies the relation $d_1 W'+d_4 W+d_5=0$,  which in turn implies that \eqref{def_second_order_equation} 
reduces to
\begin{eqnarray}
\label{eqwithcrossderivatives}
    d_1 \, \sigma'' + d_{2*} \, \ddot\sigma_*
     + d_4 \, \sigma' + d_6 \, \sigma \; = \; 0 \, \quad \text{with} \quad d_{2*}\equiv d_2+ W^2 d_1 + W d_3  \, ,
\end{eqnarray}
where $\ddot\sigma_*$ denotes the double derivative of $\sigma$ with respect to $t_*$.

Second, one introduces the radial coordinate
\begin{equation}
\label{def_rstar}
    r_*=\int \frac{\mathrm{d}r}{n(r)}\,, \qquad n =\sqrt{ - \frac{d_1}{d_{2*}}}\,,
\end{equation}
in order to get a wave equation that exhibits the usual  flat radial d'Alembertian. One must also renormalise the function $\sigma$, introducing the new function $\Psi=\sigma/{\cal N}$, where ${\cal N}$ satisfies  $d_1 ({\cal N}/{n})' + d_4 ({\cal N}/{n})=0$.

We thus obtain the wave equation
\begin{eqnarray}
\label{waveop}
    \left[\frac{\partial^2 }{\partial t_*^2} - \frac{\partial^2 }{\partial r_*^2}\right] \Psi + \Veff \, \Psi \; = \; 0 \,,
\end{eqnarray}
with the potential
\begin{eqnarray}
\label{Veffgeneral}
    \Veff = \frac{1}{d_{2*}} \left( \frac{{\cal N}''}{\cal N} d_1 + \frac{{\cal N}'}{\cal N} d_4 + d_0 \right) \, .
\end{eqnarray}

Finally, going from the time domain to the frequency domain via  a Fourier transform  with respect to the new time coordinate $t_*$, i.e. 
\begin{equation}
   \Psi(t_*,r) = \Psi(r) e^{-i\omega t_*} \, ,
    \label{Ftransform}
\end{equation}
the  equation for $\Psi$ takes the familiar Schr\"odinger-like form:
\begin{equation}
\label{Schrödinger_equation_generic}
\Schroop\Psi=-\frac{\mathrm{d}^2 \Psi}{\mathrm{d}r_*^2}+ \Veff \, \Psi \; = \; \omega^2 \, \Psi \, ,
\end{equation}
where $\Schroop$  denotes the Schr\"odinger operator and $\Veff$ should be viewed as a function of $r_*$, even if it is given explicitly only in terms of $r$.

\section{Homogeneous black holes with primary hair}
\label{sec.Primary_hair_BH}
In this section, we concentrate on the black hole solutions recently obtained for a family of  DHOST theories \cite{Bakopoulos:2023fmv} (see also \cite{Baake:2023zsq,Bakopoulos:2023sdm,Charmousis:2025xug}), within the Beyond Horndeski subclass,  characterised by the  functions\footnote{In the "beyond Horndeski" notation for the action, as used in the original papers \cite{Bakopoulos:2023fmv,Baake:2023zsq,Bakopoulos:2023sdm}, we have
    $$G_2(X)=-\frac{2 \alpha}{ \lambda^2} X^\nth,\; \quad
    G_4(X)=1- \alpha X^\nth,\; \quad F_4(X)=\frac{\alpha}{4}(2\nth-1) X^{\nth-2} \, .$$.} 
\begin{align}
\label{P_n}
    P(X)&=-\frac{2 \alpha}{ \lambda^2} X^\nth \, , \quad
    \Ftwo(X)=1- 2X A_1(X)\, , \quad
    A_1(X) =\frac{\alpha}{2} X^{\nth-1}\,, \quad 
A_3(X)=\frac{\alpha}{2}(2\nth-1) X^{\nth-2}\,,
\qquad
\end{align}
while $A_2=-A_1$, $A_4=-A_3$ and $A_5=0$.
These theories are characterised by three parameters: $\nth$, which, for simplicity, is assumed   to take positive integer or half-integer values henceforth; $\alpha$  is a dimensionless coupling constant and $\lambda$  is a coupling constant with dimension of length. 

\subsection{Black hole solutions}
These theories admit static  spherically symmetric solutions  of the form \eqref{line_element}-\eqref{phiback},
with the  scalar field profile given by
\begin{equation}
X=\frac{q^2 /2}{1+(r / \lambda)^2}\,, \qquad \psi^{\prime}(r)^2=\frac{q^2}{\cA(r)^2}\left[1-\frac{\cA(r)}{1+(r / \lambda)^2}\right] \,, \label{scalar}
\end{equation}
while the metric components read
\begin{equation}
    \cA=\cB=1-\frac{2\mu}{r}-\xi_\nth\frac{2\lambda}{r} \,\Xi_\nth(r/\lambda)\,.
    \label{solAq0}
\end{equation}
In the above expression, $\mu$ is an integration constant (with dimension of length) and  $\xi_\nth$ is a dimensionless parameter combining $q$ and $\alpha$: 
\begin{equation}
\label{xi}
    \xi_\nth\equiv \alpha ({2\nth-1}) \left({q^2}/{2}\right)^{\nth} \,.
\end{equation}
We have also introduced  the function 
\begin{equation}
   \Xi_\nth(x)\equiv\int_0^x \mathrm{d}u \; \frac{u^2}{\left(1+u^2\right)^\nth}=\frac{x^3}{3}\, {}_2F_1(3/2,p;5/2;-x^2)\,.
\end{equation}
defined via the hypergeometric function $_2F_1$.
At spatial infinity,  $\Xi_p(r/\lambda)$ tends to a constant which therefore  contributes to  the ADM mass\footnote{Note that $M$ has the dimension of length, since we work implicitly in units where $G=1$ and $c=1$.} 
\begin{equation}
\label{mass}
    M\equiv \mu+\frac{\sqrt{\pi}\,\Gamma(\nth-\frac32)}{4\, \Gamma(\nth)}\xi_\nth\, \lambda\equiv \mu+\mreg\qquad (\nth\neq -1/2, 1/2, 3/2)\,.
\end{equation}
The resulting metric describes a black hole geometry if $\cA$ vanishes for some finite radius $\reff$. Otherwise the solution describes a naked singularity (or a soliton if the metric is regular at the origin, i.e. when $\mu=0$ or, equivalently, $M=\mreg$). Let us set henceforth $\lambda$ to unity in order to simplify expressions. We will restore it if necessary later on.

\begin{figure}[!h]
\begin{center}
\includegraphics[width=8.3cm]{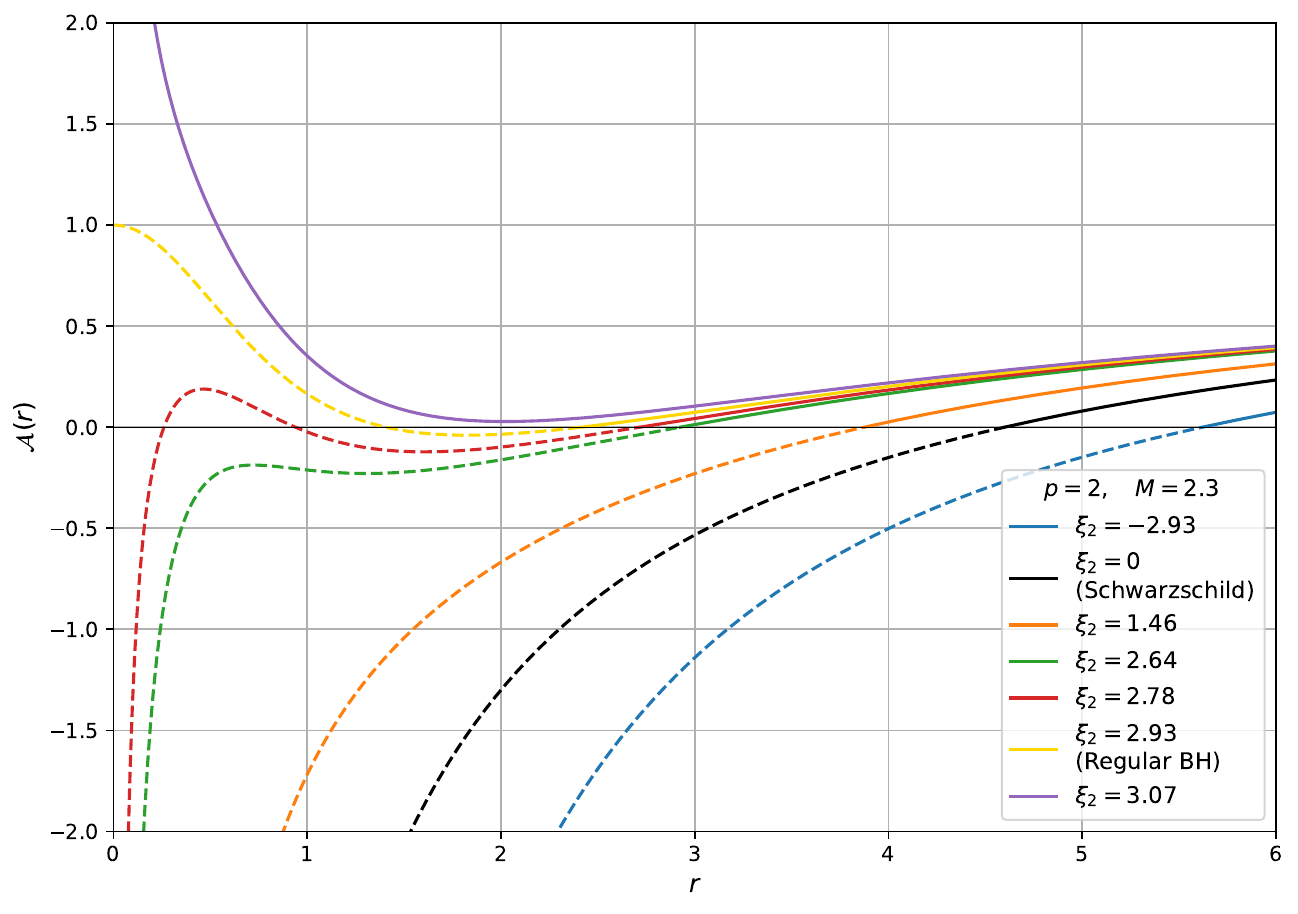}
\includegraphics[width=8.3cm]
{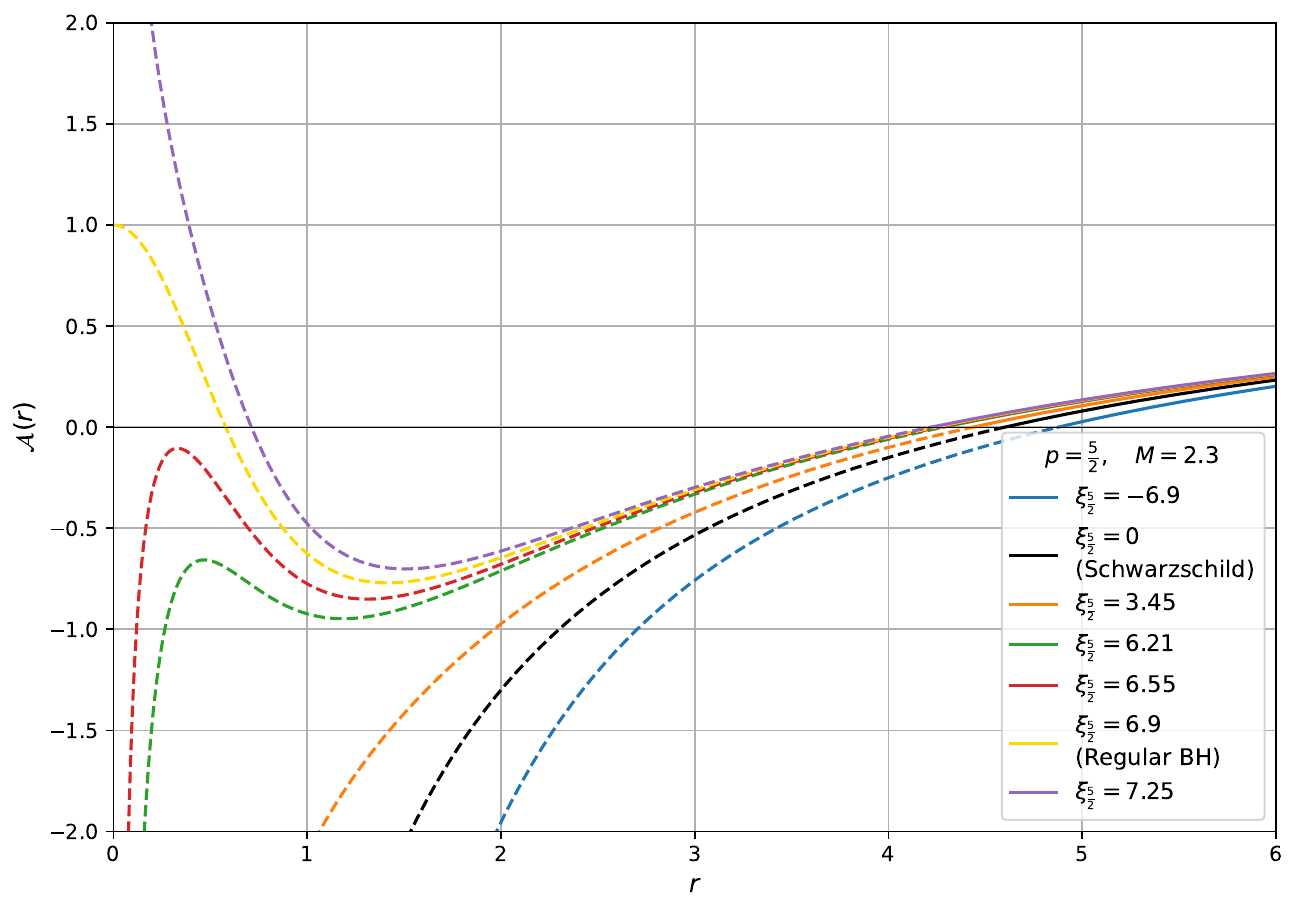}
\includegraphics[width=8.3cm]{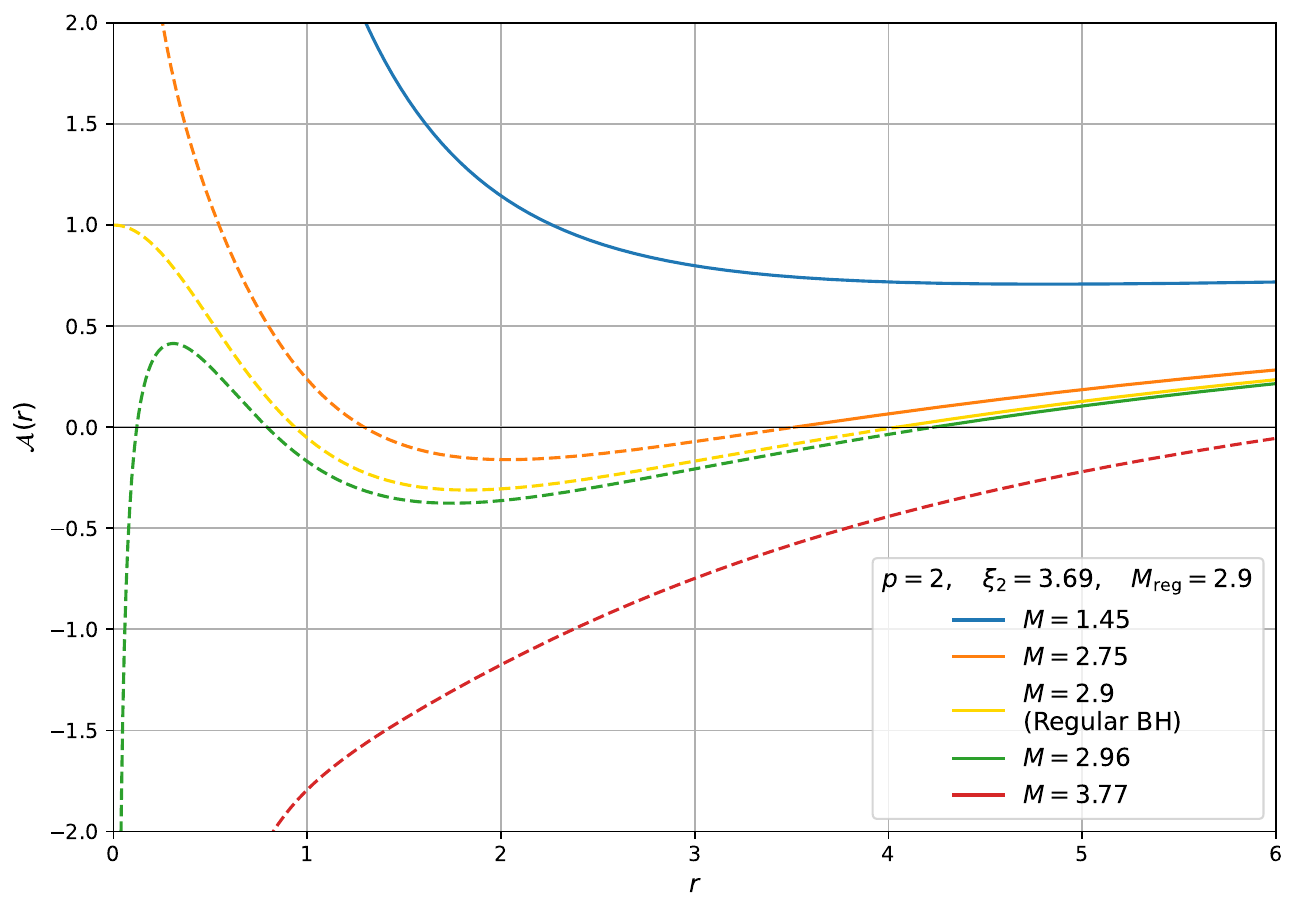}
\includegraphics[width=8.3cm]
{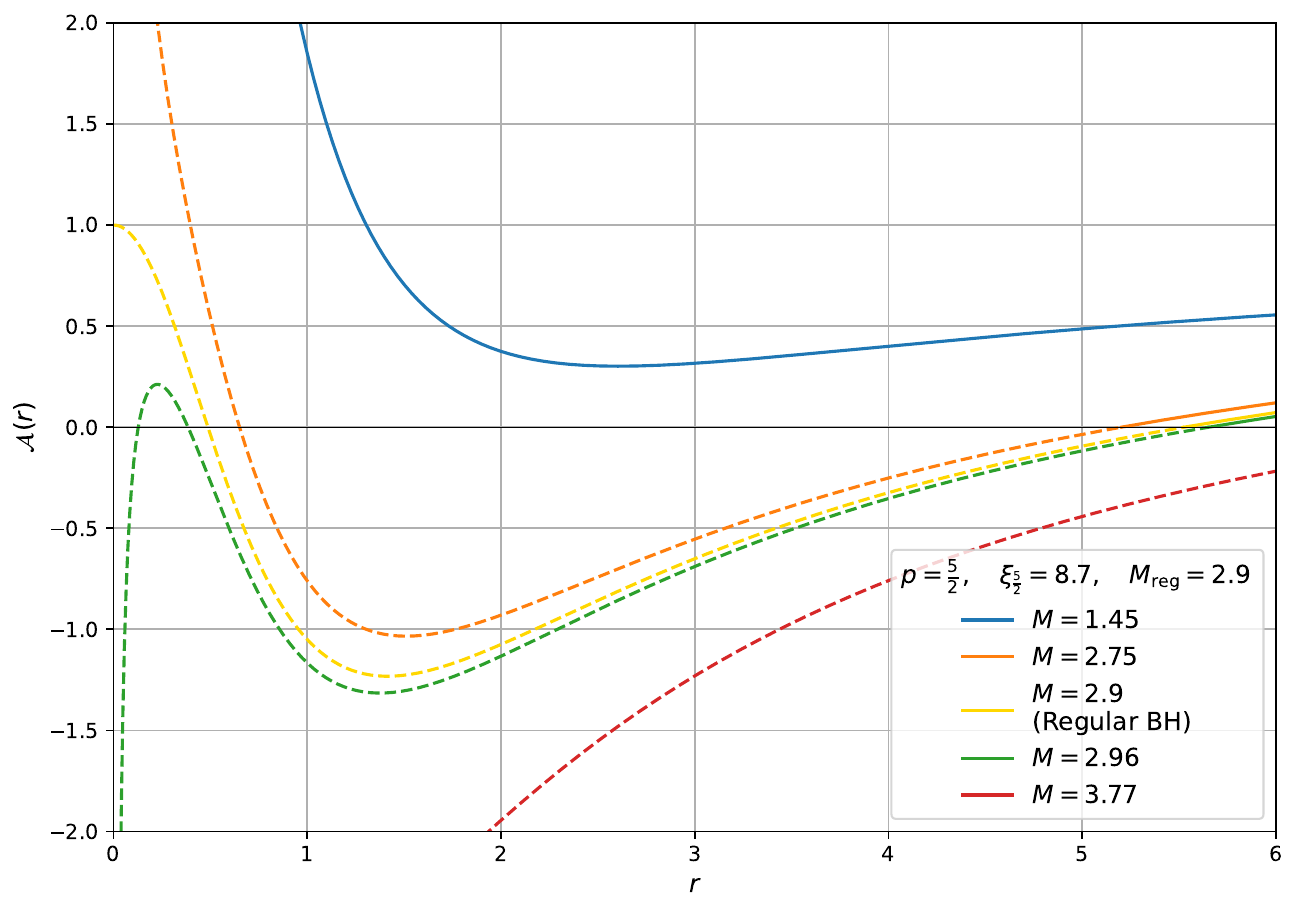}
\end{center}
\caption{\small Metric component $\cA(r)$ of the solution  in the theory  $\nth=2$ (left) and the theory $\nth=5/2$ (right). The top two graphs correspond to  $M=2$ and several values of the parameter $\xi_\nth$. The two graphs below correspond to a regular solution with $\mreg=2.9$ and several values of the mass $M$. In all these pictures,  the regions inside the horizon are indicated by a dashed curve.}
\label{Fig_A}
\end{figure}

Note that we are concentrating here on the {\it homogeneous} metrics, i.e.  for which $\cA=\cB$; the non homogeneous cases are obtained via disformal transformations and discussed later on.
Let us briefly mention a few interesting metrics associated with some particular  values of $\nth$. For $\nth=1/2$, the dimensionless parameter defined in \eqref{xi} vanishes and one obtains a stealth Schwarzschild black hole. Interestingly enough this is the only homogeneous solution belonging to Horndeski theory, since $A_3$ in \eqref{P_n} vanishes in this case.
All other values of $\nth$ involve beyond Horndeski theories within the class \eqref{P_n}. As it turns out, one has to consider non-homogeneous metrics in order to attain hairy Horndeski black holes as we will see later on. For $\nth=2$, the metric component is given by the closed-form expression
\begin{equation}
\label{A_n2}
\cA(r)=1-\frac{2 M}{r}+\xi_2\left(\frac{\pi / 2-\arctan r}{r }+\frac{1}{1+r^2}\right) \, .
\end{equation}
Another interesting case \cite{Bakopoulos:2023sdm} corresponds to $\nth=5/2$ with the metric component
\begin{equation}
\label{A_n52}
\cA(r)=1-\frac{2 M}{r}+\frac{2\xi_{_{5/2}} }{3r}\left(1-\frac{r^3}{(1+r^2)^{3/2}}\right) \,.
\end{equation}
Both solutions belong to beyond Horndeski theories \eqref{P_n} and their properties have been discussed in the literature \cite{Bakopoulos:2023fmv, Bakopoulos:2023sdm, Charmousis:2025xug}.
Several examples, corresponding to different parameters are plotted in Fig. \eqref{Fig_A}. 
In the rest of this section, we will always assume $p\geq 1$, postponing the discussion on the stealth case $p=1/2$ to Section~\ref{subsection_p=1/2}. We will also exclude the case $\nth=3/2$, which leads to a solution with no finite ADM mass. 

\subsection{Schr\"odinger-like reformulation}
We now apply the general procedure outlined in the previous section  to the above family of homogenous backgrounds for arbitrary $\nth$. 

\subsubsection{Master variable and Schrödinger equation}
As explained earlier, the system of equations can be reduced to a single second-order equation \eqref{def_second_order_equation} for a gauge-invariant master variable $\sigma$. In the present case, one finds that $\sigma$ corresponds to the combination
\begin{equation}
\label{MasterVariable_N2Beyond}
\begin{aligned}
   \mv=&  \, \xi_\nth (r^2+1)^2 \left[ \cA^2 \psi^{\prime 2}H_0 - 2 {q} \cA \psi' H_1 + q^2 H_2 + \frac{2}{q} \cA^2 \left( {\psi'^2}\delta \dot \phi - q\psi' \delta \phi'\right) - (q^2 - \cA^2 \psi'^2)(K+rK')\right]\\
   +&\, q \xi_\nth r(r^2+1)   \cA \psi' \dot K + q^2 \cA (r^2+1)^p \left[ \cA H_2 - \frac{1}{2}(3 \cA-1) K - r \cA K'\right]\, .
   \end{aligned}
\end{equation}
Let us stress that $\sigma$ is unique up to a background-dependent multiplicative factor.
Note that, in practice, since we work in the gauge \eqref{gauge}, we have obtained only the terms in $H_0$, $H_1$ and $H_2$ presented above. The full expression, valid in any gauge, can then be reconstructed by considering the gauge transformations, as discussed in Appendix~\ref{App_gauge}.

We now introduce the new time coordinate $t_*$, as defined in \eqref{def_tstar}.
In the present case, the function $W$ is simply
\begin{equation}
\label{timechange_p}
  W(r)=\frac{1}{q}\psi'(r)\,,
\end{equation}
 which means that $t_*$ coincides with the scalar field\footnote{One can check that the relation $t_*=\phiback/q$ requires the condition $(q^2-2 X)F_{X}-r^2 XP_{X}=0$
 , which is verified in the present case.}, up to a rescaling by $1/q$:
\begin{equation}
\label{tstar_p=2}
t_*=t+\frac1q\psi=\frac1q\phi\,.
\end{equation}
This is a remarkable property as the background scalar field is actually regular on the event horizon and beyond.
This signifies that although the initial ``Schwarzschild-like'' coordinates $t$ and $r$ are defined only outside the metric event horizon, here for the monopole,  the new ``Schrödinger'' time coordinate $t_*$ is extended well within the horizon interior and in many cases, as we will see, up to the central singularity at $r=0$. This can be seen explicitly  by introducing the advanced null coordinate 
\begin{equation}
\label{def_v}
    v=t+\int \frac{\mathrm{d}r}{\cA(r)}\,,
\end{equation}
in terms of which the metric becomes 
\begin{equation}
\label{line_element_v}
    \mathrm{d}s^2=-\cA \mathrm{d}v^2+2\mathrm{d}v\, \mathrm{d}r+r^2d\Omega^2\,.
\end{equation}
Similarly to the Schwarzschild metric rewritten in terms of the Eddington-Finkelstein coordinates, the above metric remains regular when  crossing any (future-directed) horizon where advanced null time is well defined. In terms of the coordinates $v$ and $r$, the new coordinate $t_*$  is  given by
\begin{eqnarray}
    t_*&=&v-\int \frac{\mathrm{d}r}{\cA}+\int \frac{\mathrm{d}r}{\cA}\sqrt{1-{\cA}/{(1+r^2)}}
    \\
    &=& v-\int \frac{\mathrm{d}r}{(1+r^2)\left(1+\sqrt{1-{\cA}/{(1+r^2)}}\right)}\,,
\end{eqnarray}
where we have combined \eqref{tstar_p=2}, \eqref{def_v} and the expression of $\psi$ inferred from \eqref{scalar}. Note that the sign ambiguity for $\psi$ in \eqref{scalar} has been resolved by requiring  the cancellation of the two divergences in the first line so as to get a regular expression in the second line. 
Moreover, the scalar field being time-like everywhere, since $\partial_\mu\phi\, \partial^\mu\phi=-2 X<0$ according to \eqref{scalar}, we deduce immediately 
that $t_*$ is a timelike coordinate everywhere (in contrast with the initial coordinate $t$ which is space-like in the local interior of the event horizon).

To obtain the Schr\"odinger form of the master equation, we must also introduce the new   radial coordinate, defined in \eqref{def_rstar}:
\begin{align}
\label{r_star_hairy}
   r_*(r)\equiv \int^r \frac{\mathrm{d}x}{n(x)}\,, \qquad  n(r)&=\frac{\left(r^2+1\right) \sqrt{r^2+1-\cA(r)}}{\sqrt{2 p-1}\ r}\,.
 \end{align}
 If the function $r^2+1-\cA(r)$ does not vanish or, equivalently, if $\mu=M-\mreg>0$ \eqref{mass}, then $n(r)$ is defined over the whole interval $]0,+\infty[$, thus not only outside the event horizon, if it exists, but also inside it,  down to the central singularity at $r=0$. In this case, the lower bound in the above integral can be chosen to be zero, and the integral converges both for $r\to 0$ and $r\to +\infty$, meaning that $r_*$ is defined on a compact interval of the form $[0,\rmax]$. If  the function $r^2+1-\cA(r)$ vanishes at some finite radius $\reff>0$, which occurs{\footnote{This can happen for black holes with an inner horizon as well as naked singularity spacetimes. Indeed, it is clear that for $M>\mreg$  we always have $r^2+1-\cA(r)>0$ since $\cA\rightarrow -\infty$ as $r\rightarrow 0$. For the regular black hole, $M=\mreg$,  we are exactly at the threshold, while only when $M<\mreg$ do we have that  $\reff^2+1-\cA(\reff)=0$ with $\reff>0$. Even then the integral is convergent up to $r=\reff$ keeping $\xi_\nth$ fixed.}} when $M<\mreg$, then $n(r)$ is defined in the interval $]\reff,+\infty[$, leading again to a compact interval for $r_*$. This case corresponds to $\mu<0$.  We first discuss the generic case $\mu>0$ and postpone this special case to Subsection~\ref{subsection_r1}.

In all cases, one can define the new coordinate $r_*$ both outside and inside the horizon, similarly to the coordinate $t_*$. Moreover, the time coordinate $t_*$ foliates spacetime in regular constant curvature slices\footnote{ It is instructive to perform the change of coordinates \eqref{tstar_p=2} in the initial background metric \eqref{line_element} with the components \eqref{solAq0}. This amounts to performing the time translation, $dt=dt_*-\frac{\psi'}{q}\mathrm{d}r$ while making use of the background solution \eqref{scalar},
\begin{equation}
 \mathrm{d} s^2 =-\cA(r) \mathrm{d} t_*^2 +\frac{2\cA \psi'}{q} \mathrm{d} t_* \mathrm{d} r+ \frac{\mathrm{d} r^2}{1+r^2} + r^2 \mathrm{d} \Omega^2  \, .
\end{equation}
The background metric is manifestly regular at the horizon given \eqref{scalar}, while interestingly constant $t_*$ hypersurfaces foliate spacetime in negative constant curvature slices. The curvature length of the constant curvature space is unity because we have set $\lambda=1$. It is given by $\lambda$ in all generality.  } while $r_*$ effectively compactifies the infinite radial distance. This has important consequences for the boundary conditions of our monopole perturbation, as we will see below.
 
\subsubsection{Effective potential}
In the coordinate chart $(t_*,r_*)$, we obtain a Schr\"odinger-like equation \eqref{Schrödinger_equation_generic} with an effective potential, which can be written (in terms of the original radial coordinate $r$) as
\begin{eqnarray}
\label{Veff_HairyBH}
    \Veff(r)&=& \frac{3 \left(r^2+1\right)^2}{16 (2p-1)(r^2+1-\cA)}\left(\frac{2\xi_\nth}{(r^2+1)^{p}}+3-\frac{(r^2+1-\cA)(5 r^2+{1})}{3r^2\left(r^2+1\right)}\right)^2\nonumber\\
    &+&\frac{p \, \xi _p}{2 p-1}\left(r^2+1\right)^{1-p}+\frac{r^2+1-\cA}{3 (2 p-1) r^4} \left(\left(3 p^2-9 p+5\right) r^4-(3p-5) r^2+2\right)\,.
\end{eqnarray}
Let us discuss the sign of this potential, recalling that we are interested in half integer and integer values of $\nth$ such that $\nth > 1/2$ and that we are focusing on the case where $r^2+1-\cA>0$. The potential \eqref{Veff_HairyBH} contains three terms: the first  is manifestly positive, the second depends only the sign of the scalar charge $\xi_\nth$, while the third is positive definite for $\nth=k/2$ with $k\in \mathbb{N}$ and $k\geq 5$. One can thus immediately conclude that  the monopole potential is positive when both $\xi_\nth>0$ and $p\geq 5/2$, whereas  the other cases 
require further inspection. 

\begin{figure}[!h]
\begin{center}
\includegraphics[width=8.3cm]{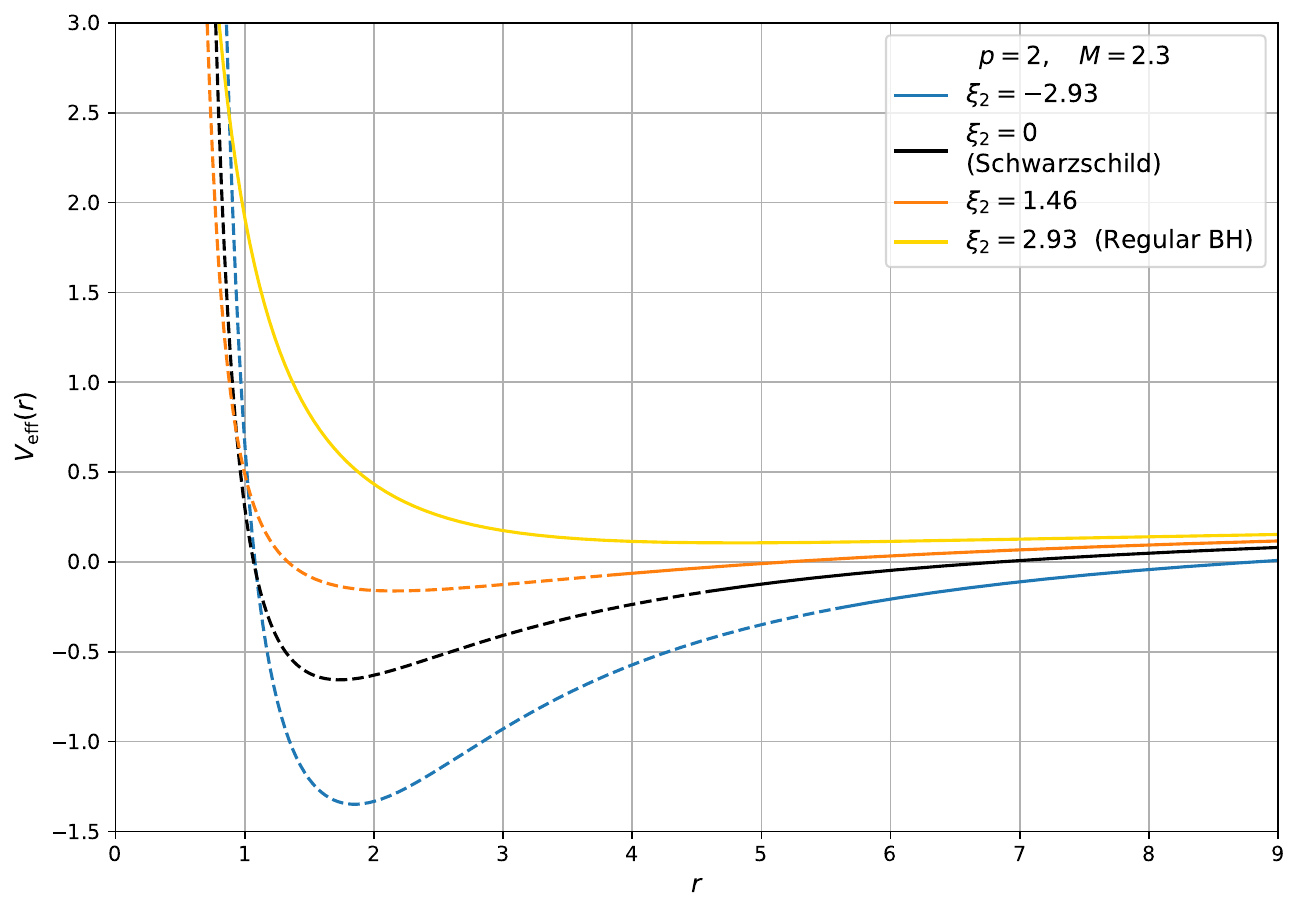}
\includegraphics[width=8.3cm]{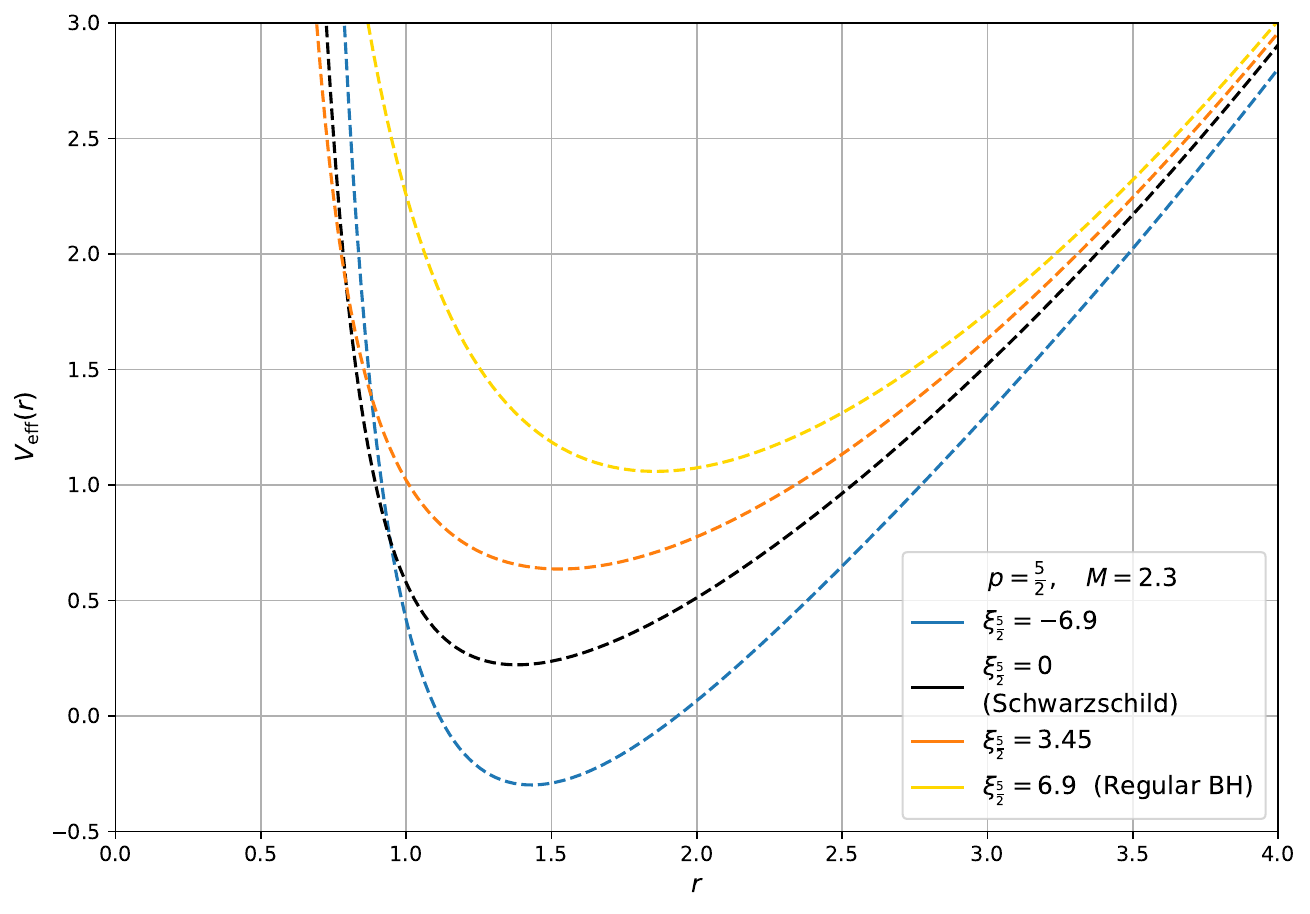}
\includegraphics[width=8.3cm]{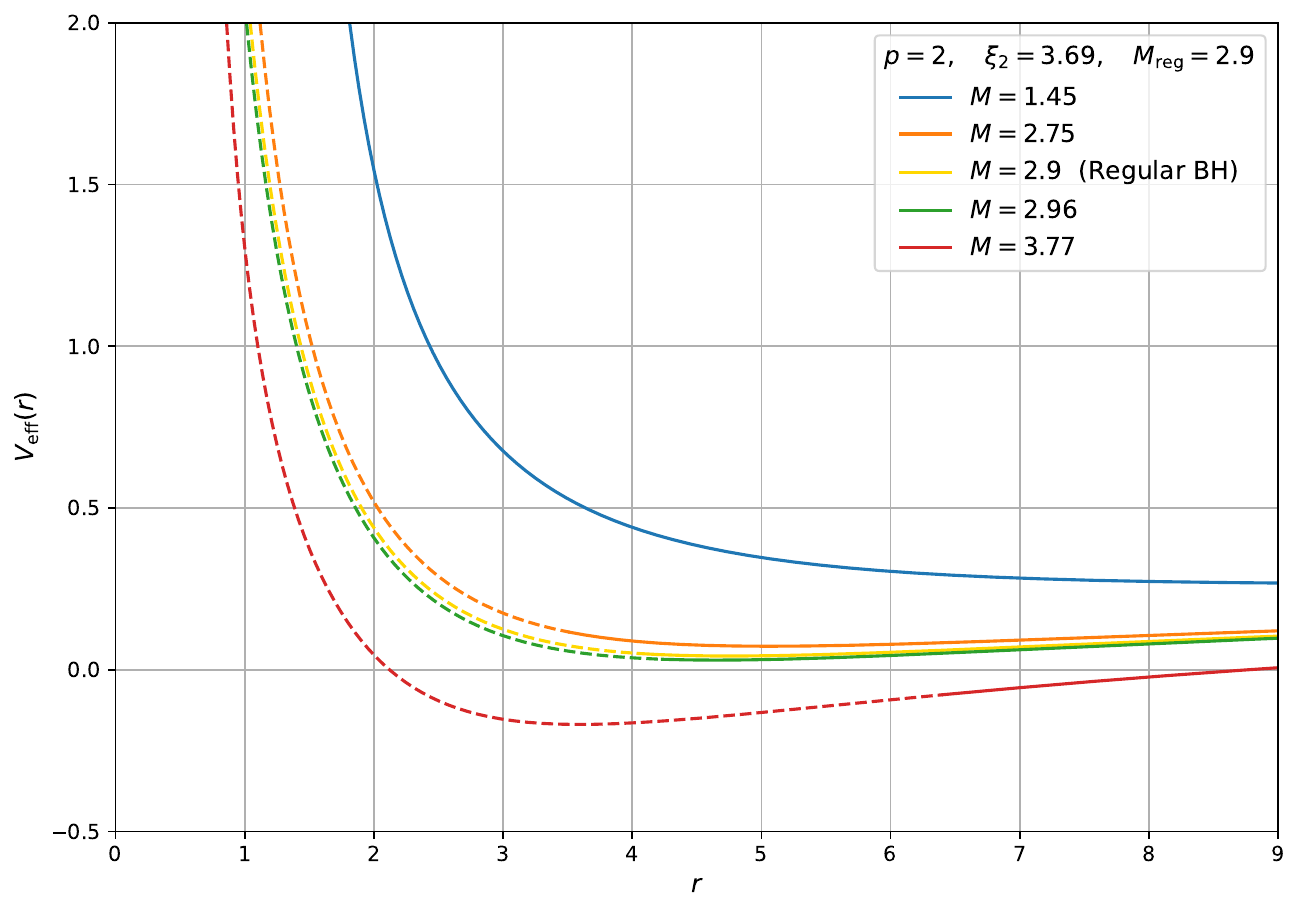}
\includegraphics[width=8.3cm]{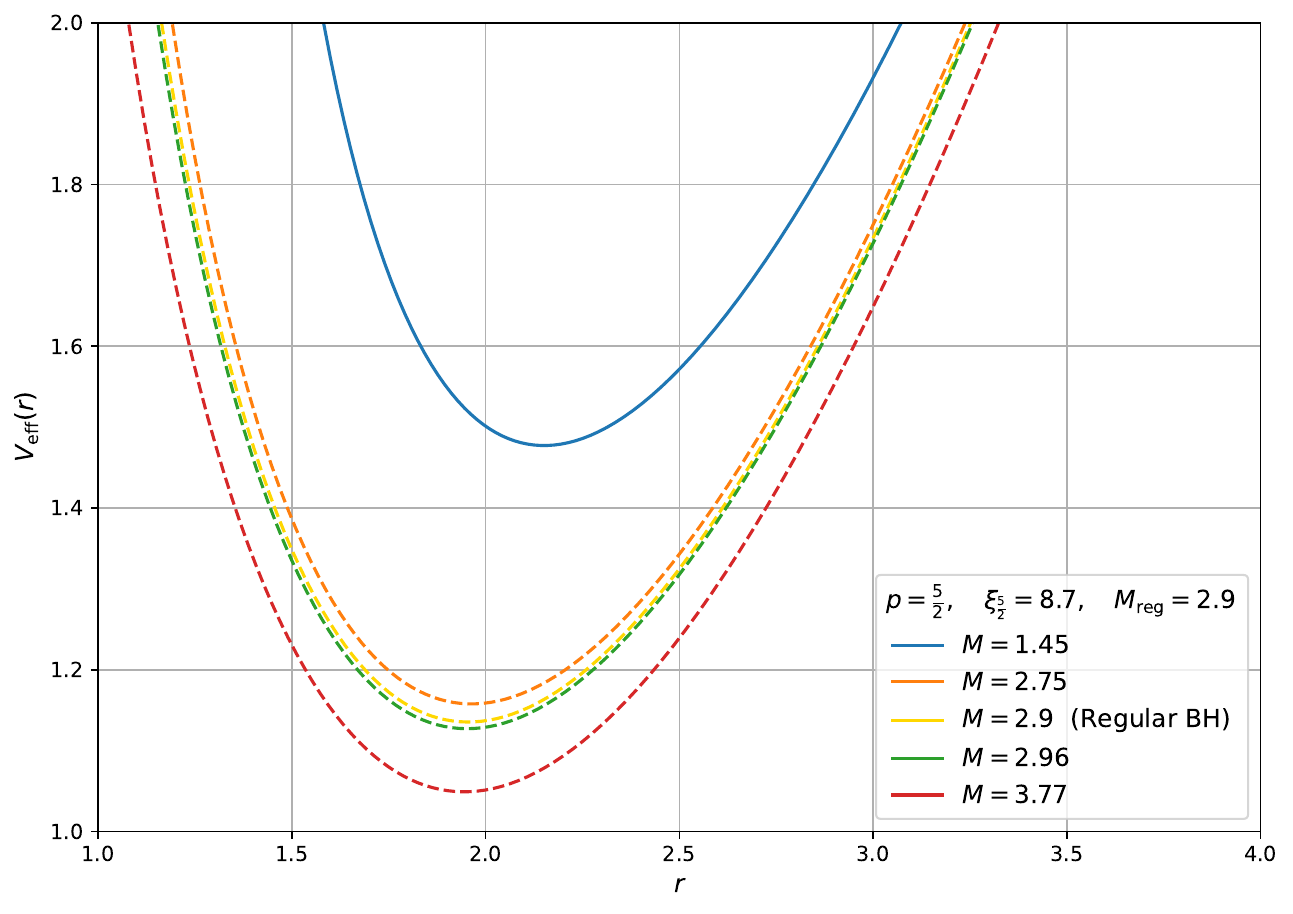}

\end{center}
\caption{\small Effective potential $\Veff$  in the theory $\nth=2$ (left) and in the theory $\nth=5/2$ (right) for several values of the parameters, corresponding to some values in Fig.~\ref{Fig_A} showing the metric component $\cA$. The top two graphs correspond to  $M=2.3$ and several values of the parameter $\xi_\nth$. The two graphs below correspond to a regular solution with $\mreg=2.9$ and several values of the mass $M$. }
\label{Fig_V}
\end{figure}

In Fig.~\ref{Fig_V}, we have plotted the  potential in the cases $\nth=2$ (left graph) and $\nth=5/2$ (right graph), for various values of the parameter $\xi_p$. When $r\to 0$, both potentials diverge, but, when $r\to\infty$, the  potential in the first case goes to a constant, whereas it goes to infinity in the second case. Note that the potential well increases in depth as we take $\xi_\nth$ to increasing negative values. 

The potential \eqref{Veff_HairyBH} can be expressed in terms of $r_*$, at least formally, by inverting the relation \eqref{r_star_hairy} between $r_*$ and $r$.
In the limit $r\to 0$, one finds that the new radial coordinate goes to zero, as
\begin{equation}
\label{asymrstar}
    r_* \; =\, \frac{\sqrt{2(2p-1)}}{5\sqrt{ \mu}} \,r^{5/2} (1+ {\cal O}(r))\,,
\end{equation}
while the potential diverges as
\begin{equation}
\label{V_limit_0}
    \Veff(r_*) \; = \; 
    \frac{11 }{100\ r_*^2} (1+ {\cal O}(r_*))\,\quad {\rm when} \quad r_*\to 0 \, .
\end{equation}

In the opposite limit, $r\to\infty$, $r_*$ tends to a finite limit, which we will call $\rmax$,  and we have 
\begin{equation}
\label{ustar}
          u_*\, \equiv \, \rmax - r_* \; = \; \frac{\sqrt{2p-1} }{r}(1 + {\cal O}(1/r)) \,.
\end{equation}
In the same limit, the 
potential \eqref{Veff_HairyBH} goes to infinity as 
\begin{equation}
\label{V_limit_max}
    \Veff\; = \, \frac{(p-2)(p-1)}{u_*^2} \left( 1 + {\cal O}(1/u_*)\right)\quad {\rm when} \quad r_*\to \rmax
    \qquad (p\neq1,2) \,, 
\end{equation}
except when $p=1$ or $p=2$.  For these two particular values, the potential instead goes to a constant,
\begin{equation}
    \Veff \to V_1^{\infty}= 2-\xi _1 \quad \text{for} \; p=1 \, ,\qquad \Veff \to V_2^{\infty}= \frac{1}{3} \quad \text{for} \; p=2 \,,
\end{equation}
as illustrated on the left plot of Fig.~\ref{Fig_V} for $\nth=2$.

In the cases where the potential \eqref{Veff_HairyBH} is not positive for all radii, we will later use the nice property that it can be expressed in the form 
\begin{equation}
\label{V_S}
    \Veff=S^2-\frac{dS}{dr_*}\,,
\end{equation}
with the explicit function\footnote{The function $S$ is defined up to an integration constant, hence other choices for $S$ are possible when solving \eqref{V_S}.}
\begin{align}
\label{Sfunction_Hairy}
S=\frac{r^2+1}{2(2p-1)\, r\, n} &\left[ \frac{\xi_p \left(r^2+1\right)^{1-p} \left(4 r^3+3 r(1-\cA)+2\mu\right)}{r(\cA-1)+2\mu} \right. \nonumber \\
&-\left.\frac{\left((5-4 p) r^2+1\right) \cA+\left(r^2+1\right) \left(4 (p-2) r^2-1\right)}{2 r^2}\right] \, ,
\end{align}
which is well defined for $r\in \, ]0,+\infty[$, and therefore for $0<r_*<\rmax$.

\subsection{Stability analysis}
We now study the stability of the monopole perturbation. This analysis is carried out in two steps: first, we examine the self-adjointness of the Schrödinger operator associated with \eqref{Schrödinger_equation_generic}; second, we analyze its positivity.

\subsubsection{Self-adjointness}
We first note that the two end points of the interval $[0,\rmax]$ are singular\footnote{Since the potential is singular at the end points.}, except for $p=1$ and $p=2$ in which case $r=\rmax$ becomes regular.
The  construction of a self-adjoint extension of the Schrödinger operator (see \cite{Kral_Zettl} for instance)  depends  on the asymptotic behaviors of the solutions at both extremities of the interval $[0,\rmax]$, which can be determined explicitly. 

In the limit $r_*\to 0$, using \eqref{V_limit_0}, 
the  solutions of the Schr\"odinger equation 
behave as 
\begin{equation}
\label{asymptr0}
\Psi=\Psi_+ + \Psi_- \, , \qquad 
\Psi_\pm\left(r_*\right) = a_\pm \, r_*^{s_\pm}\left[1+ {\cal O}(r_*)\right]\,,
\end{equation}
with $s_+=11/10$ and $s_-=-1/10$.
Both solutions $\Psi_\pm$ are square-integrable, which means that the endpoint $r_*=0$ is in the so-called limit-circle case. 

In  the limit $r_*\to \rmax$, when $p \notin \{1,2\}$ we find, using \eqref{V_limit_max},
\begin{equation}
\label{asymptr02}
\Psi=\hat\Psi_+ + \hat\Psi_- \, , \qquad 
   \hat\Psi_\pm(r_*) = b_\pm \, u_*^{s_\pm} \left[ 1 +{\cal O}(u_*) \right] \, , \quad u_*=\rmax - r_* \, ,
\end{equation}
with  $s_+=p-1$ and $s_-=2-p$. 
Only the  solution $\hat\Psi_+$, which vanishes in the limit $r_*\to \rmax$, is square integrable\footnote{The second solution is square integrable if $p<5/2$, which is excluded here since we have assumed that $p$, which is an integer or a semi-integer, does not take the values $1$, $2$ (as well as $3/2$).}. Hence, the endpoint $r_*=\rmax$ is in the limit-point case.  As a consequence, there is no freedom in the boundary condition at $r_*=\rmax$, whereas there is a one-parameter  family  of boundary conditions at the point $r_*=0$ that leads to a self-adjoint extension of the operator. These boundary conditions can be formulated from the asymptotics \eqref{asymptr0} as follows,
\begin{eqnarray}
\label{boundcondalpha}
   a_+ \, \cos \theta \,  + \,  a_- \, \sin\theta \,  = \; 0 \, ,
\end{eqnarray}
where $\theta$ is an arbitrary angle at this stage. It is straightforward to check that such boundary conditions lead indeed to a symmetric operator. If, in addition, one imposes the regularity of the solution, one obtains the so-called Friedrichs extension~\cite{Reed:1975mmp2}: one must impose  $a_-=0$ which corresponds to taking $\theta=\pi/2$ in the previous parametrisation.  

\medskip

The  cases $p=1$ and $p=2$ are special because the boundary $\rmax$ is regular\footnote{Since the potential is constant  near the boundary $\rmax$, the asymptotic solutions are  of the form 
\begin{equation}
\Psi =  e^{\pm i \kappa\,  r_*} \left( 1+ {\cal O}(u_*)\right)\,, \qquad \kappa\equiv\sqrt{\omega^2-V_{1,2}^{(\infty)}}\,,
\end{equation}
where $u_*$ was defined in \eqref{asymptr02} and $\kappa$ can be imaginary (if  $\omega^2<V_{1,2}^{\infty}$), in which case the exponentials become real.  If one expands the exponential around $\rmax$, one gets the desired asymptotic solution.}.
Hence,  since $r_*=0$ is a limit-circle end point, 
there is a 2-parameter family of boundary conditions that leads to a self-adjoint extension.  In principle, one can consider boundary conditions which mix the behaviors of the solutions at the two end-points. However, from a physical perspective, there is no justification for considering situations in which the boundary conditions at $r=0$ are coupled to those at spatial infinity.
Furthermore, we also assume  for simplicity  that $\Psi$ satisfies the same boundary condition as in the generic ($p \neq 1,2$) discussed above,
\begin{equation}
\label{boundary_p_1_2}
    \Psi(\rmax)=0\,, \qquad (p=1,2)\,.
\end{equation}
Such a condition implies that there is no perturbation  (and no radiation) at spatial infinity.

Finally, the space of square-integrable functions satisfying the chosen boundary conditions at both endpoints defines the domain of the self-adjoint Schrödinger operator. We will use the notation $\SchroopE$ for the extension of the Schrödinger operator in this domain. 
When the potential \eqref{Veff_HairyBH} is positive everywhere, one obtains immediately a positive self-adjoint operator
and one  can skip the discussion given just below and go directly to the subsection \ref{subsubmonopolemodes}.

\subsubsection{Non strictly positive potentials}

If the potential is not positive, we  use  the property \eqref{V_S} to derive,
for any $\Psi$ in the domain of the Schrödinger operator $\SchroopE$, the expression
\begin{eqnarray}
\label{psiApsi}
\int_0^{\rmax} \!\!  \mathrm{d}r_* \, \bar\Psi \SchroopE \Psi \; = \;  {\Bterm}_0 - {\Bterm}_{\rm max}  + \, 
\int_0^{\rmax} \!\!  \mathrm{d}r_* \vert \Dope\Psi \vert^2   \, ,
\end{eqnarray}
where
\begin{eqnarray}
    \Dope\Psi \equiv \partial_* \Psi + S \Psi \, ,
\end{eqnarray}
and the boundary terms are defined by
\begin{eqnarray}
\label{bounddefinition}
{\Bterm}_0= \lim_{r_* \rightarrow 0} \bar\Psi \, \Dope\Psi  \, , \qquad 
 {\Bterm}_{\rm max}=\lim_{r_* \rightarrow \rmax} \!\!\! \bar\Psi \, \Dope\Psi .
\end{eqnarray}
As a consequence, by choosing boundary conditions such that 
${\Bterm}_0 - {\Bterm}_{\rm max}>0$, one obtains a positive self-adjoint operator.

Let us start by studying the  term ${\Bterm}_{\rm max}$ in \eqref{bounddefinition}. In the cases where $p \notin \{1,2\}$, we recall that the behaviours of $\Psi$ and $S$ are respectively given by
\begin{eqnarray}
\label{psiandS_asymptot}
  \Psi(r_*)  = b_+ u_*^{p-1} \left[ 1 +{\cal O}(u_*)\right] \, , \qquad S(r_*)=\frac{2-p}{u_*} \left[ 1 +{\cal O}(u_*)\right] \,,
\end{eqnarray}
which leads to  
\begin{eqnarray}
     \overline{\Psi} \Dope\Psi \, = \, (3-2p)\vert b_+\vert^2   (u_* )^{2p-3} \left[ 1 + {\cal O}(u_*)\right]\,.  
\end{eqnarray}
In the cases  $p =1$ and $p=2$, one finds that $S$ approaches a constant as $r \rightarrow \rmax$, given respectively by
\begin{equation}
    S_1(r_*)= -\frac{\pi}{2} \left[ 1 + {\cal O}(u_*)\right]\,  \quad  \text{or} \quad    S_2(r_*)=-\frac{4 \sqrt{3}}{{3} \pi} \left[ 1 + {\cal O}(u_*)\right]\,.
\end{equation}
The boundary condition \eqref{boundary_p_1_2} thus implies that the boundary term vanishes. 
In summary, for any $p > 1/2$, the contribution of the boundary term at $\rmax$ vanishes, i.e. 
\begin{eqnarray}
    {\Bterm}_{\rm max} \, = \, 0 \, . 
\end{eqnarray}

Let us now determine the term ${\Bterm}_0$, which requires the computation of 
the sub-leading terms in the asymptotic expansion of the solution \eqref{asymptr0} and of the function $S$. 
In terms of $r$, rather than $r_*$, we find that the two independent solutions in \eqref{asymptr0} behave as follows:
\begin{eqnarray}
\label{psi_asympt}
    \Psi_+(r) &=& {c}_+ r^{11/4} \left( 
    1 - \frac{5+p}{10}r^2 - \frac{3+2\xi_p}{24 \mu} r^3+{\cal O}(r^4) \right) \nonumber \, , \\
  \Psi_-(r)   &=& {c}_
    - r^{-1/4} \left( 1+ \frac{p-1}{2} r^2 + \frac{3+2 \xi_p}{4\mu}r^3 +{\cal O}(r^4) \right) \, ,
\end{eqnarray}
while the asymptotic expansions of the functions $S$  and $n$ are respectively given by
\begin{eqnarray}
\label{S_asympt}
    S(r)= \sqrt{\frac{\mu }{4p-2} } r^{-5/2}\left( -\frac{11}{2} + \frac{4p-35}{10} r^2 - \frac{5(3+2 \xi_p)}{24 \mu} r^3 + {\cal O}(r^4)\right) 
\end{eqnarray}
and
\begin{eqnarray}
\label{n_asympt}
    n(r) = \sqrt{\frac{2\mu }{2p-1} } r^{-3/2}\left( 1+r^2 + \frac{3+2 \xi_p}{12 \mu} r^3 +{\cal O}(r^4) \right) \,.
\end{eqnarray}
As a consequence, when $r \rightarrow 0$, we obtain 
\begin{eqnarray}
     \overline{\Psi} \Dope\Psi \, = -\; \sqrt{\frac{2\mu}{2p-1} } \left[ \vert {c}_-\vert^2  \left(\frac{3 }{r^3} + \frac{9p}{5r} + \frac{7(3+2 \xi_p)}{8}\right) +3\,  {{c}_-} \overline{c}_+ \right] + {\cal O} (r) \, .
\end{eqnarray}
Hence the boundary term is well-defined only when $\Psi$ is chosen to be regular at the origin (i.e. $c_-=0$), in which case
\begin{eqnarray}
    {\Bterm}_0=0 \, .
\end{eqnarray}
In conclusion, choosing a regular boundary condition  for $\Psi$ at the origin defines a positive self-adjoint extension of our Schrödinger operator\footnote{The same approach can be generalized to other choices of the integration constant entering the definition of the $S$ function in \eqref{Sfunction_Hairy}, as discussed in Appendix \ref{sec.GeneralBoundaryConditions}.}.

\subsubsection{Positive self-adjoint operator: computing the lowest modes}
\label{subsubmonopolemodes}
Given a positive self-adjoint extension $\SchroopE$, the time evolution of $\Psi$ is formally expressed as
\begin{equation}
 \Psi(t)= \cos(\SchroopE^{1/2} t) \Psi_0 + \SchroopE^{-1/2} \sin(\SchroopE^{1/2} t) \dot{\Psi}_0 \,,
\end{equation}
where $\Psi_0$ and $\dot{\Psi}_0 $ are some smooth initial conditions with compact support~\cite{Wald:1979lth} . 
If $\SchroopE$ is positive then 
$\cos (\SchroopE^{1/2} t)$ and $\SchroopE^{-1/2} \sin(\SchroopE^{1/2} t)$  are bounded operators. Therefore the time
evolution of $\Psi$ remains bounded at all times, which guarantees the stability of the linear perturbations. 

\medskip

\begin{figure}[!h]
\begin{center}
\includegraphics[width=8.3cm]{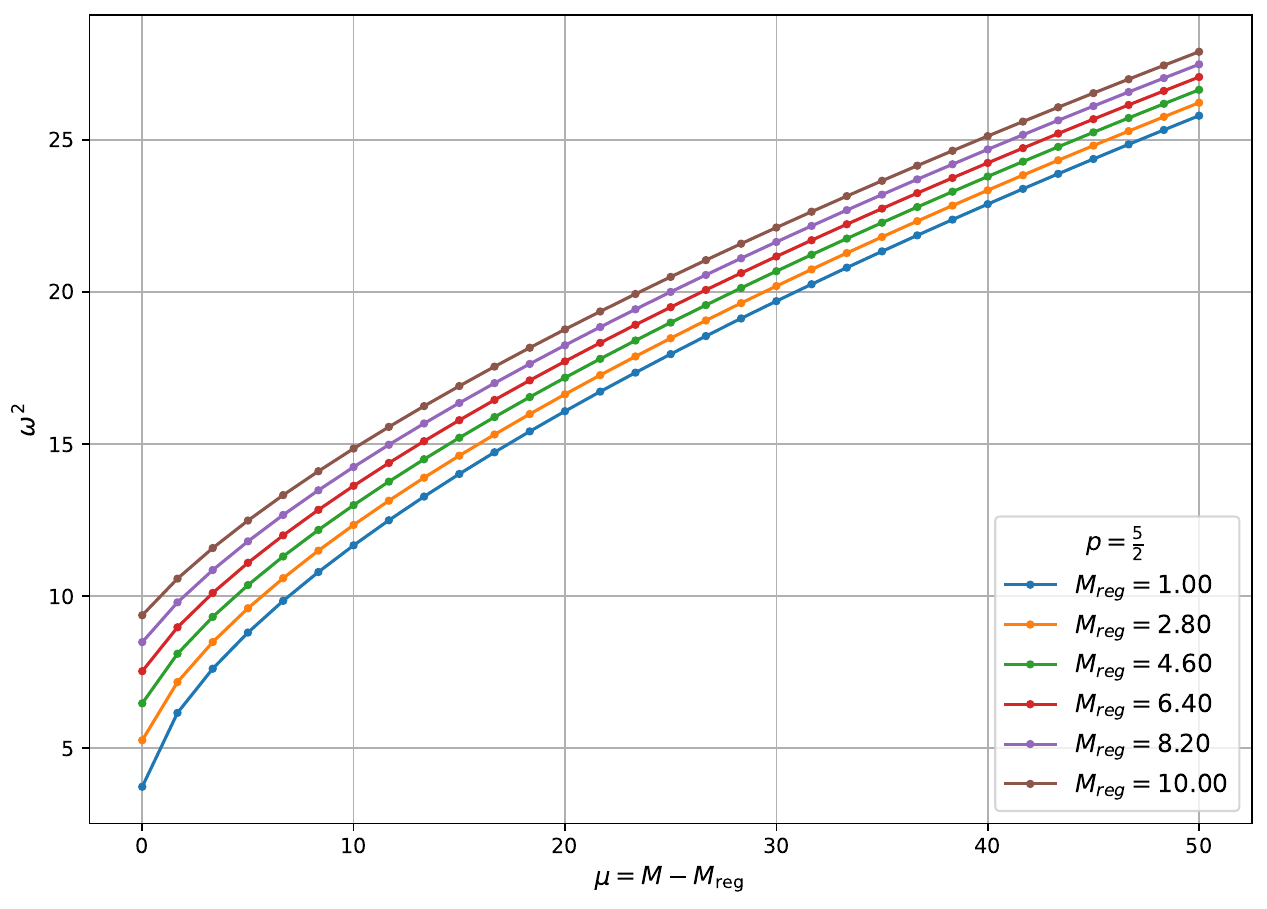}
\includegraphics[width=8.3cm]{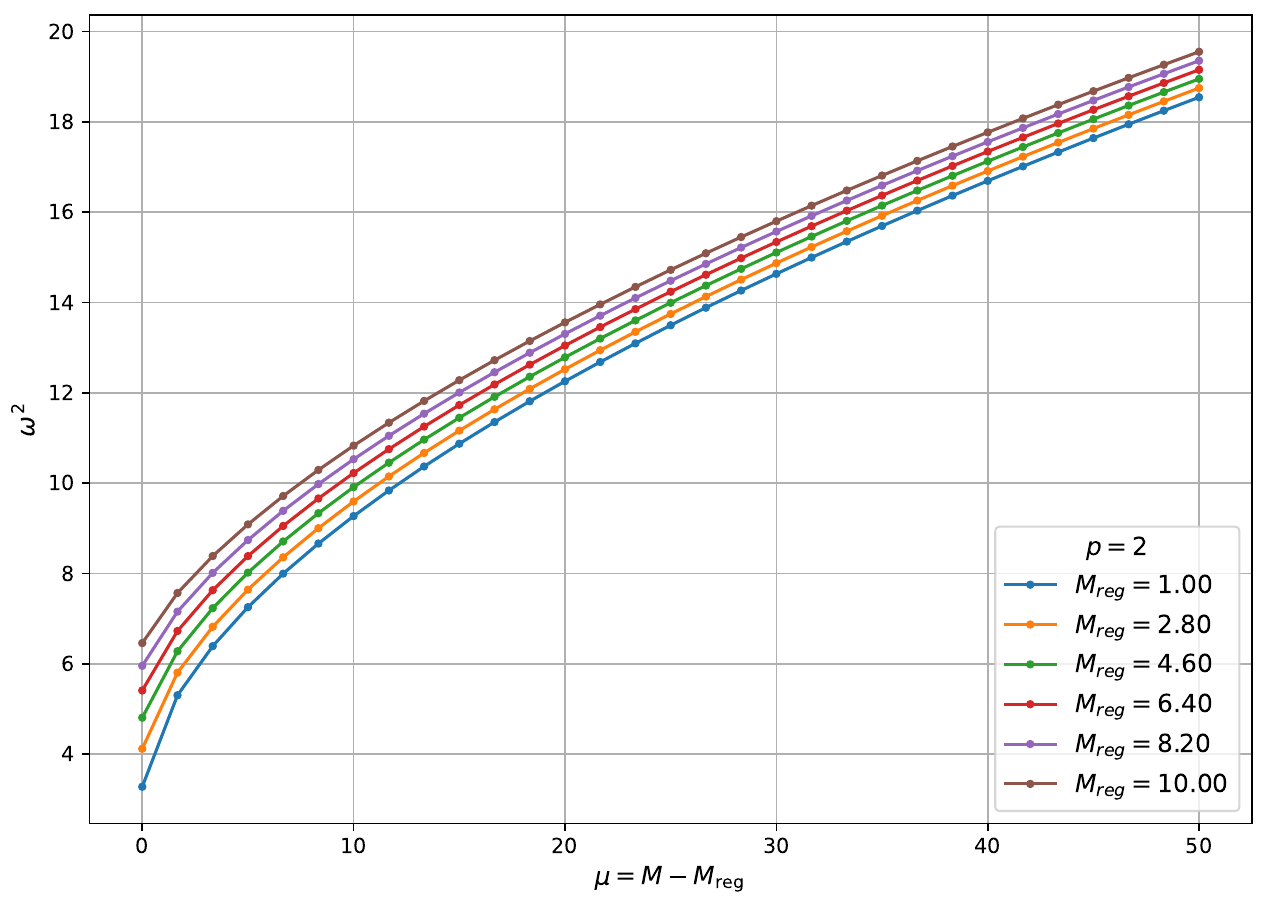}
\end{center}
\caption{\small{ These plots show the square of the first (fundamental) normal mode in terms of $\mu$ for different values of $\mreg$ (see \eqref{mass}) for $p=5/2$ (left) and $p=2$ (right). Note that  we have set $\lambda=1$.}}
\label{Omega0withM0}
\end{figure}

Radial stability can be confirmed by an explicit  numerical computation of the fundamental mode frequency. Indeed, as illustrated in Fig.~\ref{Omega0withM0}, computing the numerical values of the lowest mode frequency for different values of the parameters, we see that  $\omega^2>0$.

\subsection{Case $\mu\leq0$}
\label{subsection_r1}
Let us now consider the case $\mu\leq0$, which implies that there exists $\reff\geq0$ such that 
\begin{eqnarray}
\reff^2+1-\cA(\reff)=0 \, .    
\end{eqnarray} 
As mentioned earlier, this can happen for black holes with an inner horizon as well as naked singularity spacetimes and we redirect the reader to Fig.~4.2 of \cite{Charmousis:2025xug} for a detailed ``phase diagram'' in the $p=2$ case. For all the cases here, when we switch off the scalar charge $\xi_\nth=0$, we get a naked singularity with negative ADM  mass $\mu$.
Recall that the coordinate $r_*$ still lives in a compact interval with the only difference that the lower bound  now corresponds to $r=\reff$.

The asymptotic limit at spatial infinity is the same as in the case $\mu>0$ so we only need to examine  the lower bound of the interval.
For $\mu<0$, the Schrödinger radial coordinate behaves as 
\begin{equation}
        r_* \, \propto \, \sqrt{u} \left(1+ {\cal O}(u) \right)\,, \quad u\equiv r-\reff
\end{equation}
so that 
\begin{eqnarray}
\label{asymptSu0}
    \Veff=\frac{3}{4 r_*^2}\left(1+ {\cal O} (r_*)\right) \qquad \text{ and   } \qquad 
    S&=&\frac{1}{2 r_*}\left(1+ {\cal O} (r_*)\right) \, .
\end{eqnarray}
Hence, the solution to the Schrödinger equation when $r\to \reff$ behaves as
\begin{equation}
\label{asymptu0}
\Psi=\Psi_+ + \Psi_- \, , \qquad 
\Psi_\pm\left(r_*\right) = a_\pm \, r_*^{s_\pm}\left[1+ {\cal O}(r_*)\right]\,,
\end{equation}
with $s_{+}=3/2,s_{-}=-1/2$. Since $\reff$ is a singular point and only $\Psi_+$ is square-integrable, we are in the limit-point case and the operator does not need an additional boundary condition to be extended to a self-adjoint operator.

\medskip

Let us now consider the case where $\mu=0$ which corresponds to a regular compact object. In this very special case, $\reff=0$ and when $r\to 0$,
\begin{equation}
        r_* \, \propto \, r \left(1+ {\cal O}(r) \right)
\end{equation}
Similarly to what have been done in the case $\mu <0$: 
\begin{eqnarray}
\label{asymptS_mu0}
    \Veff=\frac{2}{ r_*^2}\left(1+ {\cal O} (r_*)\right) \qquad \text{ and   } \qquad 
    S&=&\frac{1}{ r_*}\left(1+ {\cal O} (r_*)\right) \, .
\end{eqnarray}
Hence, one can solve the Schrödinger equation when $r\to \reff$ and get
\begin{equation}
\label{asymptu0Mu0}
\Psi=\Psi_+ + \Psi_- \, , \qquad 
\Psi_\pm\left(r_*\right) = a_\pm \, r_*^{s_\pm}\left[1+ {\cal O}(r_*)\right]\,,
\end{equation}
with $s_{+}=2,s_{-}=-1$. Since $\reff=0$ is a singular point and only $\Psi_+$ is square-integrable, we are in the limit-point case and the operator does not need an additional boundary condition to be extended to a self-adjoint operator. 

\medskip

In both cases, we can use the asymptotics \eqref{asymptu0} and \eqref{asymptu0Mu0} to show that the boundary terms  in \eqref{bounddefinition} vanish, ensuring radial stability.

\medskip

\section{Quadratic Horndeski theories}
\label{section_Horndeski}

In this section, we focus on quadratic Horndeski theories. We start with a quick review of the equations of motion for a static and spherically symmetric metric \eqref{line_element} associated with a scalar field of the form \eqref{phiback} in generic Horndeski theories. Then, we study some important properties of the equations of the monopole around such a solution. Finally, we examine the dynamics and the stability of the monopole in different background solutions of Horndeski theories.

\subsection{Equations: background and perturbations}
\label{Strat Monop}

In \cite{Bakopoulos:2023fmv}, it was shown that, within a general shift- (and parity-) symmetric beyond Horndeski theory, the background equations for  a static, spherically symmetric metric ansatz \eqref{line_element} and the scalar field ansatz \eqref{phiback} reduce to three simple independent equations,
\begin{eqnarray}
\label{easy}
 &&  r^2(P Z)_{X}+2(\Ftwo Z)_X \left(1-\frac{q^2 \gamma^2}{2 Z^2 X}\right)=0, \qquad \frac{\cA}{\cB}=\frac{\gamma^2}{Z^2}, \\
\label{easy3}
 &&   2\gamma^2 \left(\cA r-\frac{q^2 r}{2 X}\right)'=-r^2 P Z-2\Ftwo Z\left(1-\frac{q^2 \gamma^2}{2 Z^2 X}\right)+\frac{q^2\gamma^2X' r}{ZX^2}\left( 2X\Ftwo_{X}-\Ftwo\right),
\end{eqnarray}
where $\gamma$ is an integration constant and the function $Z$ has been defined by
\begin{equation}
\label{Z}
    Z=-\Ftwo-2 X A_1 \, .
\end{equation}
The first equation in \eqref{easy} enables us to find $X$ as a function of $r$; the second equation in \eqref{easy} gives the relation between $\cA$ and $\cB$; the remaining equation \eqref{easy3} enables to determine $\cA$. 
Let us note that, in Horndeski theories, the condition $Z_X = 0$ which ensures the homogeneity of the metric, strongly restricts the space of theories since the unique free function $F$ is constrained to be of the form \eqref{P_n} with $p=1/2$, in which case there is a stealth solution. Hence, for Horndeski theories, we have in general $Z_X \neq 0$ which leads to non-homogeneous black hole solutions, i.e. with $\cA\neq\cB$.

\medskip
From now on, we assume that there is a solution of the system \eqref{easy}-\eqref{easy3}. In order to find the Schrödinger-like equation for the monopole perturbation around this background, we follow the strategy described in Section \ref{perturbations section}. 
Hence, we start with the three equations \eqref{tt_equation}, \eqref{tr_equation} and \eqref{rr_equation} for the three unknowns $H_0$, $H_1$ and $H_2$. In the case where $X$ is not constant, we can easily implement the background equations to simplify these perturbation equations and we do so by replacing all the occurrences of $\cB$, $P(X)$ and their derivatives by the expressions given by the background equations \eqref{easy} and \eqref{easy3}.
The case where $X$ is a constant can be treated following the same procedure and a theory where $X= q^2/2$ will be treated as a special case in Section \ref{stealth section}.

After a direct calculation, we find the master variable  \eqref{nnvv}, the two algebraic equations    \eqref{EqE1} and \eqref{EqE2}, the associated determinant $\Det$ \eqref{determinant_def} which can be seen not to vanish in general. All this ensures that $\sigma$ satisfies a second order partial differential equation. 

Finally, to pass from the second order equation to the Schrödinger one, we change the time coordinate according to \eqref{def_tstar}, we compute the Schrödinger coordinate $r_*$ and we deduce the effective potential $\Veff$ \eqref{Veffgeneral}. We provide some details in the Appendix \ref{Moredetails}, however the general expressions for the potential and the Schrödinger coordinate are too lengthy to be included in this paper. We therefore provide a GitHub page where the full expressions for all these quantities can be downloaded in their most general form \cite{Iteanu2026rph}.

\subsection{Disformed  black hole  with  primary hair}
In this subection, we illustrate the results of the previous subsection by considering Horndeski theories derived from the Beyond Horndeski theories \eqref{P_n} via a disformal transformation. Indeed, as discussed explicitly in Appendix \ref{sec.ConfDis}, it is possible to perform a  disformal transformation \eqref{disformal_transf} of the original metric, i.e.
\begin{eqnarray}
    \tilde{g}_{\mu\nu} = C(X) \, g_{\mu\nu} + D(X) \, \partial_\mu\phi \, \partial_\nu\phi \,,
\end{eqnarray}
to map one of the theories \eqref{P_n} governed by the action $S[g_{\mu\nu},\phi]$  into a Horndeski theory governed by the new action $\tilde S$ such that 
\begin{equation}
    \tilde S[\tilde g_{\mu\nu},\phi] \, = \, S [g_{\mu\nu},\phi] \, .
\end{equation}
The disformal transformation is assumed to be invertible\footnote{When the disformal transformation is not invertible, the two actions are not equivalent even in vacuum and the disformal transformation can introduce new degrees of freedom as in mimetic gravity \cite{Langlois:2018jdg}.}. As the theories \eqref{P_n} already belong to the beyond Horndeski class, the conformal factor is trivial, i.e. $C(X)=1$. 

For a general value of $\nth$, both the disformal transformation and the resulting Horndeski theory  can be defined  only implicitly. As an illustration, we discuss below two cases in which the Horndeski theory is explicitly defined:  the case $\nth=2$ where the disformal transformation can be computed; the  trivial case $\nth=1/2$ where the original theory is already of Horndeski type.

\subsubsection{Disformed theory for $\nth=2$}
For $\nth=2$, the disformal transformation that maps the theory \eqref{P_n} into a Horndeski theory is given explicitly by
\begin{equation}
    \label{D_H_p2}
    C(X)=1 \, ,\qquad 
    \cDis(X)=\frac{3\alpha}{2} X\,.
\end{equation}
This transformation has been studied in detail in Appendix B of \cite{Charmousis:2025xug} and we summarize the main results here. Let us first recall that it
preserves the signature of the metric (i.e. $g$ and $\tilde{g}$ have the same signature) in the region of  space-time where $1-3\alpha X^2 >0$, which corresponds to $\xi_2<1$, and we restrict our analysis to this region.

We can easily compute the expression of the functions that enter in the Horndeski action $\tilde{S}$. After a direct calculation, see Appendix \ref{Detailsofp2toHorn}, we find two non-trivial functions entering in the quadratic part of the Horndeski action
\begin{align}
\label{def_N0H}
    &
    \tilde P(\XH)= \frac{\sqrt{2}}{3} \frac{1-\Upsilon }{\sqrt{\Upsilon +1}} \, ,\qquad
    \tilde{F}(\XH)= \frac{\sqrt{2}}{3}\frac{\Upsilon +2}{\sqrt{\Upsilon +1}} \, , \qquad
\tilde{A}_1(\XH)= -\frac{\sqrt{6\alpha}}{3} \frac{\sqrt{\Upsilon -1}}{\Upsilon +1}\, ,
\end{align}
where we have introduced the notation
\begin{eqnarray}
    \Upsilon \equiv \sqrt{1 + 12 \alpha \tilde{X}^2} \,,
\end{eqnarray}
and  assumed $\alpha >0$ for simplicity. The case $\alpha<0$ is very similar and is discussed in \cite{Charmousis:2025xug}.

The background solution in the Horndeski frame can also be obtained explicitly. As we show in  Appendix \ref{sec.ConfDis}, the background metric can be brought into a diagonal form after the following change of the time variable
\begin{equation}
    t \; \mapsto \; \tilde t = t + \tau(r)\qquad \text{with} \qquad \tau(r) = q \int \frac{D \, \psi'}{ \mathcal{A} - D q^2} \, \mathrm{d} r \,.
\end{equation}
Nonetheless, the metric is no longer homogeneous and its components are now given by
\begin{align}
&  \tilde\cA(r)=\cA(r)-\frac{\xi_2}{1+r^2} =1-\frac{2M}{r}+\xi_2\frac{\pi/2 -  \arctan(r)}{ r}\, ,\\
&  \tilde \cB(r)= \frac{(1+r^2)^2}{(1+r^2)^2-\xi_2}\, \tilde \cA(r) \, ,
\end{align}
where we have used the definition \eqref{xi} of $\xi_2$.

As for the scalar field, its expression is obviously unchanged but it is helpful to write it in the coordinate system where the Horndeski metric is diagonal, i.e.
\begin{eqnarray}
    \phi = q \, \tilde t + \tilde\psi(r) \, \qquad \text{where} \quad \tilde{\psi}(r)=\psi(r) - q \tau(r) \, .
\end{eqnarray}
This amounts to a transformation of the radial component of the scalar field. Furthermore, $\tilde \psi$ can be obtained  from the expression of $\XH$,
\begin{equation}
\XH(r)=\frac{q^2(1+r^2)}{2\left[(1+r^2)^2-\xi_2\right]}\,,
\end{equation}
combined with those of $\tilde\cA$ and $\tilde\cB$.

Now, we have all the ingredients to compute the Schrödinger-like equation for the monopole. First of all, using \eqref{W_Horn}, we can compute the Schrödinger time coordinate $\tilde t_*$ \eqref{def_tstar} in the Horndeski frame and, interestingly, we find that it coincides with the Schrödinger  time coordinate in the original beyond Horndeski frame, since
\begin{equation}
    \tilde t_* = \tilde t + \tilde \psi / q = t_* \, .
\end{equation}
Furthermore, the radial Schrödinger coordinate $\tilde r_*$ is also equal to the original Schrödinger coordinate $r_*$. Indeed, using 
the definition \eqref{def_rstar}, one gets
\begin{eqnarray}
    \tilde n(r) &= & \frac{{\left(r^2+1\right)^{1/2} \left[-\left(r^2+1\right) \tilde \cA(r)-\xi_2 +r^4+2 r^2+1\right]^{1/2}}}{\sqrt{3} r} \nonumber \\ &=&\frac{{\left(r^2+1\right) \left(r^2+1-\cA(r)\right)^{1/2}}}{\sqrt{3} r}
    =n(r) \, .
 \end{eqnarray}
Finally, the effective potential obtained using  \eqref{Veffgeneral} coincides with \eqref{Veff_HairyBH}. 

In summary, we find that, not only the  dynamics of the monopole degree of freedom is identical for the disformally related solutions and theories, but also that the associated ``Schrödinger'' coordinates are identical.

\subsubsection{Horndeski theory  $\nth=1/2$ }
\label{subsection_p=1/2}
Within the class of theories \eqref{P_n},
the particular case $\nth = 1/2$ corresponds to a Horndeski theory, characterised by the functions
\begin{equation}
P(X)=-2 \alpha \sqrt{X}, \quad F(X)=1-\alpha \sqrt{X}, \qquad A_1(X)=-F_X=\frac{\alpha}{2 \sqrt{X}} \, .
    \label{n=1/2}
\end{equation}
It admits a stealth solution \cite{Bakopoulos:2023fmv} for which the metric coincides with the Schwarzschild solution, i.e.  
\begin{eqnarray}
\label{Schwarmetric}
    \cA(r)=\cB(r)=1-\frac{2M}{r} \, ,
\end{eqnarray}
even if the scalar field profile is non trivial and can be computed from the expression,
\begin{eqnarray}
    X(r)=\frac{  q^2}{2\left(1+r^2\right)} \, .
\end{eqnarray}

In this particular case, one cannot apply the strategy described in Section \ref{Strat Monop} to analyze the monopole, as  the determinant 
\eqref{determinant_def} (see \eqref{D_Horndeski}) vanishes.
However, in this specific case, one can sum the equations \eqref{tt_equation} and \eqref{tr_equation} so that the resulting equation  only contains the variable $H_2$ and its derivatives. One can then reabsorb the $\dot H_2$ term by redefining the time coordinate through
\begin{eqnarray}
\label{unitary}
    t_*=t+\psi(r)/q \,.
\end{eqnarray}
Then, in this new system of coordinates, the perturbation equations simplify drastically and  reduce  to the following set of three equations\footnote{
As a point of comparison, we also recall the equations for the monopole in General Relativity where the gauge we used is no longer complete. Thus, if we pick up the gauge $H_1=K=0$, we get,
\begin{equation}
    H_2+r \cA H_2'=0 \, ,\qquad \dot{H_2}=0, \qquad  H_2+r \cA H_0'=0 \, ,
\end{equation}
which leads to $H_2=1/(r \cA)$ with no time dependency, contrary to the solution here \eqref{solforH2}.}
\begin{eqnarray}
     &&{H_2} + r \cA \, H_2^{\prime} = 0 \, , \label{H2eq}\\
      &&  H_0^{\prime}=H_2'- {\varepsilon}\frac{ r \cA }{n}\frac{\partial^2 H_2}{\partial t_*^2}  \, ,\label{H0eq}\\
      && H_1= \frac{q r {\cA}^2}{\sqrt{2} \alpha  \left(r^2+1\right)^{5/2} (q^2 + X\cA)}\frac{\partial{H_2}}{\partial t_*}+\frac{\cA \psi ' }{2 q}H_0+\frac{q }{2 \cA \psi '}H_2 \, ,\label{H1eq}
\end{eqnarray}
where $\varepsilon \in \{-1,+1\}$ is the sign of $\alpha q$ while the function $n(r)$ in \eqref{H0eq} is given by\
\begin{eqnarray}
    n(r)=\frac{\vert \alpha q \vert}{\sqrt{2}}\left(r^2+1\right)^{3/2} \left(1 +   r^2 -  \cA \right)\, = \, \frac{\vert \alpha q \vert}{r \sqrt{2}}\left(r^2+1\right)^{3/2} \left( r^2 + 2M \right) \, .
\end{eqnarray}
We are using the  notation $n(r)$ as this function  plays the same role as in \eqref{def_rstar} if we define the new radial coordinate $r_*$ as
\begin{eqnarray}
    r_*(r) \; = \; \int \frac{\mathrm{d}r}{n(r)} = \frac{\sqrt{2}}{\vert \alpha q \vert} \int \frac{\mathrm{d}r}{r\left(r^2+1\right)^{3/2} \left(r^2+ 2M \right)} \, .
\end{eqnarray}

The first equation \eqref{H2eq} can be easily integrated and the general solution for $H_2$ depends on an arbitrary function $f(t_*)$ according to
\begin{eqnarray}
\label{solforH2}
  H_2(t_*,r)=\frac{f(t_*)}{r\cA(r)} = \frac{f(t_*)}{r-2M} \, .
\end{eqnarray}
The second equation \eqref{H0eq} can also be integrated and its general solution depends on another arbitrary function $g(t_*)$ according to
\begin{eqnarray}
    H_0(t_*,r)=g(t_*) + \frac{f(t_*)}{r-2M} + r_*(r) \, \frac{\mathrm{d}^2 f}{\mathrm{d} t_*^2} \, .
\end{eqnarray}
Finally, the remaining equation \eqref{H1eq} fixes uniquely $H_1$. 
We thus find that the solution to the perturbations equations is very peculiar, since the radial profile is imposed by the equations. 
We conclude that there is no monopolar degree of freedom at the linear level in this particular case.

\subsection{Stealth solution with a constant kinetic density $X$}
\label{stealth section}
This subsection is devoted to the study of the monopole in another class of stealth solutions in Horndeski theories where the kinetic density is constant.  

\subsubsection{Background theory and equations}
A sufficient condition for a DHOST theory to admit a stealth Schwarzschild  solution  \eqref{Schwarmetric} is to impose that the effective stress–energy tensor vanishes\footnote{In generic DHOST theories (where the tensor sector is not degenerate), we can write the equations of motion for the metric as $G_{\mu\nu}= T_{\mu\nu}^{\rm eff}$ where $G_{\mu\nu}$ is the Einstein tensor and $T_{\mu\nu}^{\rm eff}$ is an effective stress-energy tensor.}. This was analyzed in \cite{Minamitsuji:2019shy} and revisited in \cite{Langlois:2021aji} for instance. When restricting  to Horndeski theories,  stealth solutions with a scalar field of the form \eqref{phiback}  are obtained if the conditions
\begin{eqnarray}
\label{stealth_X0_conditions}
    X=X_0= q^2/2, \quad P\left(X_0\right)=P_X\left(X_0\right)=Q_X\left(X_0\right)=0 \,
\end{eqnarray}
are satisfied, hence $X$ is necessarily a constant $X_0$. In that case, the scalar field satisfies
\begin{eqnarray}
\label{stealthfiedl}
    \phi=qt+\psi(r) \qquad \text{with} \qquad \psi'(r)^2 = \frac{{ 2q^2Mr}}{(r-2M)^2} \, .
\end{eqnarray}
Furthermore, as shown in \cite{Langlois:2021aji},  one can restrict, without loss of generality, the analysis of linear perturbations to a subclass of Horndeski theories defined by the functions
\begin{equation}
\begin{aligned}
& F(X) \equiv 1+\alpha\left(q^2-2X\right)+\frac{\beta}{2}\left(q^2-2X\right)^2 \,, \\
& P(X) \equiv \frac{\zeta}{2}\left(q^2-2X\right)^2, \quad Q(X) \equiv \frac{\delta}{2}\left(q^2-2X\right)^2 \, ,
\end{aligned}
\end{equation}
where  $\alpha$, $\beta$, $\zeta$ and $\delta$ are constants ($\alpha$ here should not be confused with the constant introduced in \eqref{P_n}). These functions 
 satisfy the above conditions \eqref{stealth_X0_conditions} 
and  any higher order term  in powers of $\left(q^2-2X\right)$ would be irrelevant. Note also that the previous parameter $q$ now appears in the theory functions and is thus fixed.

\subsubsection{Monopole equations}
Even if the monopole in such theories has already been studied in the literature \cite{deRham:2019gha,Khoury:2020aya,Takahashi:2021bml}, it is instructive to revisit it in our approach.
Although the results derived for a generic theory in Section \ref{Strat Monop} assumed   that $X \neq q^2 / 2$ in order to implement the background equations of motion, the study of the monopole about the stealth solution with $X=q^2 / 2$ can still be carried out in a similar manner. 
In particular, starting from the monopole equations, one can exhibit a master variable \eqref{nnvv} and two equations of the form \eqref{EqE1} and \eqref{EqE2}. Hence, we can still define the determinant \eqref{determinant_def} which, for this theory, is given by
\begin{eqnarray}
 \Det=   \frac{8   q^8 \zeta \left(\alpha ^2+\beta \right)  \left(2 \alpha  q^2+1\right)^2 \left(4 \alpha  q^2-4 \beta  q^4+1\right)}{r^2 \left[ \left(4 \alpha  q^2-4 \beta  q^4+1\right)\cA +4 \beta  q^4 -2 \alpha  q^2\right]^4} 
 {(1-\cA) \cA^3}\, .
\end{eqnarray}
One immediately observes that the determinant vanishes whenever only one of the coupling constants among $\alpha,\beta,\zeta$ is non-zero\footnote{The determinant  vanishes as well  when the spacetime is flat $\cA=1$, which indicates that the flat space-time is strongly coupled in these theories.}.  This observation is consistent with the analysis of the monopole in the so-called ``decoupling limit'' that was carried out in the Appendix G of \cite{Roussille:2023sdr}, and  it suggests that, in such cases, the monopole has no dynamics. Thus, from now on, we will assume that $\zeta \neq 0$ and  $\alpha^2+\beta\neq 0$.

\medskip

We now construct the coordinate system $(t_*,r_*)$ where the second order equation for the master variable takes a Schrödinger-like form. To this end, it is convenient to introduce the length scale $\reff$ defined by its square according to
\begin{eqnarray}
    \reff^2=4\frac{ (\alpha ^2+\beta) (4 \alpha  q^2-4 \beta  q^4+1)}{\zeta  \left(2 \alpha  q^2+1\right)^2} \, .
\end{eqnarray}
At this stage, $\reff$ is not necessarily real but we assume this is indeed the case (with $\reff>0$) and we will shortly discuss the case where $\reff^2<0$ later on. 
Then, we construct the radial Schrödinger coordinate $r_*$ \eqref{def_rstar} through the function $n$,
\begin{eqnarray}
\label{nstealth}
    n(r) \; = \; \frac{(r^2-\reff^2)\sqrt{1-\cA(r)}}{\reff r} = {\frac{\rsc^{1/2}(r^2-\reff^2)}{\reff r^{3/2} }} \, ,
\end{eqnarray}
where we have introduced the notation $\rsc={2M}$ for the Schwarzschild radius. 
Interestingly, $\reff$ does not depend on the mass of the background metric but it is completely fixed depending only on the parameters of the theory.  The expression of $r_*$ (up to an irrelevant integration constant) follows immediately from \eqref{def_rstar} which gives
\begin{eqnarray}
     r_*(r)=\frac{\reff^{3/2}}{2{\rsc}^{1/2}} \left[{4 \sqrt{\frac{r}{\reff}} + \log \left(\frac{ \sqrt{r} -\sqrt{\reff}}{\sqrt{r} +\sqrt{\reff}}\right)-2 \arctan \sqrt{\frac{r}{\reff}} } \right]  \, .
\end{eqnarray}
Similarly to the tortoise coordinate in General Relativity, the Schrödinger radial coordinate belongs to the whole real line.

Next, we construct the time coordinate  \eqref{def_tstar}  through
\begin{equation}
     W(r)= \frac{1}{q}\psi'(r) + \frac{r^{5/2}}{\rsc^{1/2}(r^2-\reff^2)} \, .
\end{equation}
This expression depends on the choice of the branch for $\psi$ \eqref{stealthfiedl} and we choose the positive one,
\begin{equation}
  \psi(r)=+q\int \frac{\sqrt{1-\cA(r)}}{\cA(r)} \, \mathrm{d}r \, = \, q \left[ 2 \sqrt{\rsc r} - \rsc \ln \left(\frac{\sqrt{r}+\sqrt{\rsc}}{\sqrt{r}-\sqrt{\rsc}}\right)  \right] + \text{cste}\, ,
\end{equation}
so that $t_*$ is regular at the horizon. Indeed, if we introduce the null coordinate $v$ \eqref{def_v}
in terms of which the metric becomes \eqref{line_element_v}, the coordinate $t_*$ can be written as follows
\begin{eqnarray}
    t_*=
     v+ 2 \sqrt{\rsc r} - r - 4M \ln \left( \sqrt{r} + \sqrt{\rsc}\right) \, ,
\end{eqnarray}
which makes obvious that $t_*$ is regular at the horizon\footnote{The choice of the other branch for $\psi$ leads to a divergence when $t_*$ is expressed in terms of the ingoing variable $v$ but it is well-defined when written in term of the outgoing coordinate $u$.}.

We finally compute the effective potential,
\begin{equation}
\label{Potentiel_Stealth}
\Veff(r)=\frac{\rsc \left(3 r^2-11 \reff^2\right) \left(r^2-\reff^2\right)}{16 \reff^2 r^5} \, .
\end{equation}
Thus, we have all the ingredients entering in the Schrödinger-like equation for the monopole.  
Notice that the equation becomes elliptic for theories with $\reff^2<0$, hence there is no wave propagation and we discard this case.

\subsubsection{Stability analysis}
 To prove the stability of the monopole, we follow the same approach as previously, applying the S-deformation method. 
For that, it is convenient to introduce the dimensionless radial coordinate $x=r/r_s$ and to express the functions $n$ \eqref{nstealth} and $\Veff$ \eqref{Potentiel_Stealth} in terms of $x$ as follows
\begin{eqnarray}
    n(r) = \frac{\rsc^{1/2}}{\reff^{1/2}} \, \nu(x) \, ,\qquad   \Veff(r) =  \frac{\rsc}{\reff^{3}} \, V(x) \, ,
\end{eqnarray}
where the dimensionless functions $\nu$ and $V$ are
\begin{eqnarray}
     \nu(x)=\frac{x^2-1}{x^{3/2}} \, ,\qquad  \quad V(x) = \frac{(3x^2-11)(x^2-1)}{16 x^5} \, .
\end{eqnarray}
Then, the  Schrödinger-like equation for the monopole can also be reformulated in terms of $x$ as follows,
    \begin{eqnarray}
    -\frac{\mathrm{d}^2 \Psi}{\mathrm{d} x_*^2} + V(x) \Psi(x) \; = \; \Omega^2 \, \Psi(x) \,,
\end{eqnarray} 
where the new Schrödinger coordinate  $x_*$ (more precisely its differential $\mathrm{d}x_*$) and the dimensionless frequency $\Omega$ are now defined by
\begin{eqnarray}
    \mathrm{d}x_* = \frac{\mathrm{d}x}{\nu(x)} \, , \qquad 
    \Omega = \frac{\reff^{3/2}}{\rsc^{1/2}}\omega \, .
\end{eqnarray}
At this stage, it is easy to see that $V(x)$ can be written in terms of an explicit S-function, i.e.
\begin{eqnarray}
    V = S^2 - \frac{d S}{d x_*} \quad \text{with} \quad S(x) = \frac{x^2-1}{4 x^{5/2}} \, .
\end{eqnarray}
Since $S(x)$ is regular for $x \in ]1,+\infty[$ and $\lim_{x \to \infty} S=\lim_{x \to 1} S=0$, then, following \cite{Ishibashi:2003ap,Kimura:2017uor,Kimura:2018eiv} for instance,  the Schrödinger operator associated to the monopole is positive and the monopole is stable.

\section{Conclusion}
We have investigated the radial oscillations of black hole solutions within the framework of  DHOST theories. Our main objective was to assess  the radial stability of several static black hole solutions reported  in the literature, particularly the recently discovered solutions with primary hair \cite{Bakopoulos:2023fmv,Baake:2023zsq,Bakopoulos:2023sdm, Charmousis:2025xug}, but earlier solutions too \cite{Minamitsuji:2019shy, Charmousis:2021npl}.

We have first outlined a general procedure to identify the single physical degree of freedom --- a gauge-invariant quantity mixing scalar and metric perturbations --- and derived the equation of motion that it obeys. By introducing appropriate new time and radial coordinates, we have shown that the equations of motion can be recast in a Schr\"odinger-like form in the frequency domain. In the context of GR, our approach is analogous to the standard method  leading to the Regge-Wheeler equation for the axial modes \cite{Regge:1957td} and the Zerilli equation for polar modes \cite{Zerilli:1970se}. However, in modified gravity, the new radial coordinate no longer coincides with the usual tortoise coordinate of GR; furthermore, the  time coordinate  also needs to be modified upon going to a Schr\"odinger like form for the monopole equation.

We applied our procedure to the black hole solutions with primary hair \cite{Bakopoulos:2023fmv,Baake:2023zsq,Bakopoulos:2023sdm}, which represent a one-parameter deformations of Schwarzschild parametrised by the scalar charge. Remarkably, for this particular family of solutions,  the new ``Schr\"odinger'' time coordinate coincides with the scalar field itself --- this is not the case for other beyond-GR solutions that we examined. Another distinctive feature is that the new ``Schr\"odinger'' radial coordinate covers the whole spacetime, 
from spatial infinity down to the BH central singularity, while spanning a bounded interval.
This contrasts with GR where the tortoise coordinate spans the entire real line but covers only the exterior geometry of the black hole, limited by the horizon. 

The resulting Schr\"odinger-like equation, defined on a finite interval, is characterized by its potential. We found that this potential diverges at both boundaries of the interval (i.e. at the singularity and at spatial infinity), except in the specific cases of  theories $\nth=1$ and $\nth=2$, where it remains  finite at spatial infinity. We studied the boundary conditions for these singular Sturm-Liouville problems and showed the  existence of a positive self-adjoint operator extension, ensuring the radial stability of the solution. Additionally, we numerically computed  the fundamental mode frequencies in illustrative cases and analyzed their dependence on the BH's mass and scalar charge.   

Interestingly, we explicitly checked that the Schr\"odinger-like equation obtained for one solution remains the same for all solutions that are connected  to the original metric via invertible conformal-disformal transformations.  Thus, even if  two related solutions, associated with distinct theories, describe different physics (assuming ordinary matter is minimally coupled to each metric\footnote{In particular, their causal structure differs, since it is related to the geodesic propagation of light.}), their purely gravitational sectors are identical. Moreover, the corresponding new time and radial coordinates for the Schrödinger monopole equation coincide. In other words the monopole perturbations of different gravitational backgrounds are identical modulo an arbitrary disformal/conformal transformation of our gravitational sector.

Our results on the linear radial  stability of BHs with primary hair complement our recent work on axial perturbations around the same solutions \cite{Charmousis:2025xug}. To complete the linear analysis, 
it remains to  study the non-radial polar perturbations, which we leave for future work. This task is more complex in DHOST theories since the polar modes  involve two degrees of freedom-instead of one in GR --- due to the presence of the background scalar field.

In this work, we  also applied  the same formalism to stealth BH solutions, where the metric is Schwarzschild despite a non-trivial scalar field profile. A notable example is the stealth solution with a non-constant  kinetic term $X$ \cite{Bakopoulos:2023fmv}, obtained within the same family of theories that yield BHs with primary hair. We  showed  that the radial perturbations of this stealth BH do not yield a propagating degree of freedom at the linear level.
According to our findings the same holds for the monopole sector of any conformal/disformal class obtained starting from the stealth metric.
Interestingly, we found a similar conclusion
for radial perturbations of the BH solution in 4d Einstein-Gauss-Bonnet theory \cite{Charmousis:2021npl} when applying our formalism. 
It would be interesting to extend these analyses to the non linear level and investigate whether there exists a weaker version of Birkhoff's theorem valid for these theories.

In addition, we revisited prior stealth solutions \cite{Minamitsuji:2019shy} identified in the literature, this time with a constant $X$. At the level of linear perturbations, the corresponding DHOST theories can be characterised by just four constant parameters. Except for special values of these parameters --- where no propagating mode exists --- one generally finds  a propagating degree of freedom\footnote{It is intriguing  however that in the zero mass flat spacetime limit we find a vanishing determinant $\Delta$.}. Surprisingly,  however, the Schr\"odinger radial coordinate now spans the entire real line (as the tortoise coordinate does in GR) but  it now covers the region outside an effective horizon that does not coincide with the horizon of the metric. This effective horizon depends on the scalar charge and the theory parameters so could be either smaller or larger than the Schwarzschild radius. Again, we find that the mode is stable.  

In summary, for most cases studied in this work, we identified   a stable propagating degree of freedom. Intriguingly, this mode is not restricted to live outside the horizon, as in GR, which could lead to  puzzling phenomenological consequences. 
It would also be interesting to see if our linear results could be extended to the nonlinear time and space dependent exact solutions. This may be suggested from the presence of bound modes found here.

\subsubsection*{Acknowledgments}This work was partially supported by the French National Research Agency (ANR) via Grant No. ANR-22-CE31-0015-01 associated with the project Strong. CC thanks Richard Britto, Pedro G S Fernandes and Adrien Kuntz for interesting discussions.

\newpage
\begin{appendices}
\section{Gauge fixing and master variable}
\label{App_gauge}
In this appendix, we review some basic properties of gauge fixing in black hole perturbation theory. Details can be found in \cite{Kobayashi:2014wsa} for instance.

\subsection{Transformation of perturbations}
\label{Sec:Transfo_pertubations}
Under an infinitesimal diffeomorphism
$x^\mu\rightarrow x^\mu+\xi^\mu$,
the metric perturbations and the scalar perturbation transform as 
\begin{equation}
h_{\mu\nu}\rightarrow h_{\mu\nu}
-(\nabla_\mu\xi_\nu+\nabla_\nu\xi_\mu )\, ,
\qquad
\delta\phi\rightarrow \delta\phi
-\xi^\mu\partial_\mu\phiback,
\end{equation}
For monopole perturbations \eqref{pertgeneral} about the background \eqref{line_element} and \eqref{phiback}, the  non-trivial components of the vector field $\xi$ are restricted to be of the form,
\begin{equation}
\xi^t= T(t, r) , \qquad \xi^r= R(t, r) \, .
\end{equation}
 Furthermore, the  non-trivial components of the monopole perturbations can be parametrized as follows:
\begin{align}
&h_{t t}  =\cA(r)  H_0(t, r) \, ,\quad
h_{t r} = H_1(t, r) \, , \quad 
h_{r r}  =\frac{1}{\cA(r)}  H_2(t, r) \, , \\
&h_{a b}  = K(t, r) g_{a b}=r^2 K\gamma_{ab} \,, \quad
\delta \phi= \delta \phi(t, r) \, ,
\end{align}
where $g_{a b}=r^2 \gamma_{ab}$ and $\gamma_{ab}$ is the metric on the unit-2-sphere, with $a,b \in \{\theta,\varphi \}$.

Therefore, gauge transformations on the non-trivial components of the monopole perturbations are explicitly given by
\begin{align}
\label{H0_gauge_transfo}
 H_0(t, r)  &\rightarrow \tilde H_0(t, r) = H_0(t, r)+ 2 \dot{T}(t, r)+\frac{A^{\prime}(r)}{\cA(r)} R(t, r), \\
 H_1(t, r)  & \rightarrow \tilde H_1(t, r)=H_1(t, r)+\cA(r) T^{\prime}(t, r)-\frac{\dot{R}(t, r)}{B(r)}, \\
 H_2(t, r)  & \rightarrow \tilde H_2(t, r) = H_2(t, r)+\cA(r)\frac{R(t, r) B^{\prime}(r)-2 B(r) R^{  \prime}(t, r)}{B(r)^2},\\
K(t, r)  & \rightarrow \tilde K(t, r) = K(t, r)-\frac{2}{r} R(t, r), \\ 
\label{deltaphi_gauge_transfo}
\delta \phi(t,r)  &\rightarrow  \delta \tilde \phi(t,r) = \delta \phi(t,r) -q T(t,r) -R(t,r) \psi'(r) \, .
 \end{align}

\subsection{Gauge invariant perturbations}
It is useful to construct gauge invariant perturbations in order to relate  different gauge choices. 
For that purpose, we first express the gauge functions $R$ and $T$ in terms of the perturbations before and after the gauge transformations as follows:
 \begin{eqnarray}
R &= &\frac{r}{2}(K-\tilde{K}) \, ,\\
\label{Gauge_Fixing_T}
T &= & \frac{1}{q}\left[\delta{\phi}-\delta\tilde\phi+\frac{r}{2}(\tilde{K}-K) \psi^{\prime}\right] \, .
\end{eqnarray}
Substituting these expressions into \eqref{H0_gauge_transfo}-\eqref{deltaphi_gauge_transfo}, one finds three independent invariant perturbations:
\begin{align}
&\label{Horndeski_gauge_invariant1} 
I_0=H_0+\frac{2 \overset{.}{\delta \phi }}{q}+\frac{r \cA' K}{2 \cA}-\frac{r \overset{.}{K} \psi '}{q} \, , \\
&\label{Horndeski_gauge_invariant2} 
I_1=H_1-\frac{\cA K \left(r \psi ''+\psi '\right)}{2 q}-\frac{r \overset{.}{K}}{2 \cB}-\frac{r \cA K' \psi '}{2 q}+\frac{\cA \delta \phi '}{q} \, , \\
&\label{Horndeski_gauge_invariant3} I_2=H_2+\frac{\cA \left(r \cB'-2 \cB\right) K}{2 \cB^2}-\frac{r \cA K'}{\cB}\, .
\end{align}
The master variable $\mv$ that encodes the single propagating degree of freedom of the monopole sector can be written as a linear combination of these gauge invariant variables \eqref{Horndeski_gauge_invariant1}–\eqref{Horndeski_gauge_invariant3}. 
Its expression, originally derived in the gauge $K=\delta\phi=0$, is recovered by combining \eqref{Horndeski_gauge_invariant1}–\eqref{Horndeski_gauge_invariant3} in such a way that the resulting variable reproduces the same dependence of $\mv$ on the metric perturbations $H_0$, $H_1$, and $H_2$.

\section{Miscellaneous  on the dynamics of perturbations in Horndeski
}
In this section, we collect additional useful results concerning the dynamics of perturbations.

\subsection{On the perturbation equations}
\label{App:Perturbation_Equation}
In Section \ref{perturbations section}, we have introduced the equations for the metric perturbations $\mathscr{E}_{\mu\nu}$ for $\mu,\nu \in \{t,r,\theta,\varphi\}$.  These equations are defined from the quadratic action for the perturbations \eqref{pertgeneral} as follows:
\begin{equation}
\mathscr{E}_{\mu\nu}[h,\delta \phi] = \frac{1}{\sqrt{-{\bar{g}}}}\frac{\delta S^{(2)}}{\delta h^{\mu\nu}} \qquad \text{with} \qquad 
    S^{(2)}[h_{\mu\nu},\delta \phi] = \int d^4x\,\sqrt{-\bar{g}}\;\mathcal{L}^{(2)} \, ,
\end{equation}
where $\mathcal{L}^{(2)}$ corresponds to the quadratic term in the expansion of the Lagrangian in the perturbations. 

\medskip

Because of the symmetries of the background solution,  only 5 out of the 10 equations are non-trivial. Moreover,  spherical symmetry implies that $\mathscr{E}_{\theta\theta}$ and $\mathscr{E}_{\varphi\varphi}$ are equivalent, and more precisely $\mathscr{E}_{\theta}^{\theta}=\mathscr{E}_{\varphi}^{\varphi}$ where indices are raised or lowered with the background metric. Hence, we have at most 4 remaining independent equations. 

Furthermore, one can show that for a generic Horndeski solution,  $\mathscr{E}_{\theta \theta}$ is a linear combination of $\mathscr{E}_{t t}, \mathscr{E}_{r r}, \mathscr{E}_{t r} $ and their time and spatial derivatives,
\begin{equation}
\begin{aligned}
   \mathscr{E}_{\theta \theta}&=\frac{r^3 \cA^{\prime}}{4  \cA^2} \mathscr{E}_{t t}+\frac{\gamma ^2 r^2 \psi ' \left(r \cA Z_X X'-Z \left(r  \cA^{\prime}+2 \cA\right)\right)}{2 q Z^3}\mathscr{E}_{t r} \\
   &+\frac{\gamma^2 r^2\left(3 r Z \cA^{\prime}+4  \cA\left(Z-r X^{\prime} Z_X\right)\right)}{4 Z^3}\mathscr{E}_{r r}\\
   &-\frac{r^3 \gamma^2  \cA \psi^{\prime}}{2 q Z^2}\partial_r\mathscr{E}_{t r}+\frac{r^3 \gamma^2 \cA}{2 Z^2}\partial_r\mathscr{E}_{r r}+\frac{r^3  \psi'}{2 q\cA}\partial_t\mathscr{E}_{t t}
   -\frac{r^3 }{2 \cA}\partial_t\mathscr{E}_{t r} \, .
\end{aligned}
\end{equation}

\medskip

One might wonder also about the perturbed equation of the scalar field $\mathscr{E}_{\delta \phi}$. From  the Bianchi identity \cite{Babichev:2015rva}, we can easily show that
\begin{equation}
 \mathscr{E}_{\delta \phi} \propto \nabla^t \mathscr{E}_{t t}+\nabla^r \mathscr{E}_{r t} \, .
\end{equation}
As expected, it is not an independent equation.

\subsection{More details for the monopole in Horndeski theories}
\label{Moredetails}
In this subsection, we provide more details of the calculation presented in the subsection \ref{Strat Monop} concerning the dynamics of the monopole in Horndeski theories.

First of all,  a direct calculation shows  that the master variable \eqref{nnvv} is given, up to an irrelevant  normalization function, by the following combination\footnote{When $\cA=\cB$ or equivalently when $Z=-\gamma$, the master variable  reduces to $H_2$ in the gauge we have chosen here. Furthermore, in that case, the equations \eqref{tt_equation}-\eqref{rr_equation} drastically simplify since many of their coefficients vanish.}:
\begin{eqnarray}
\label{mv_Horndeski}
    \mv&=&\frac{2 \gamma ^2 \cA  \Ynew }{r Z^3}H_1+\frac{\gamma ^4 \cA^2   \Znew}{r Z^6}H_2
    +\frac{{\Ynew}^2}{r   \Znew}H_0 \nonumber \\
    &-& \frac{Z\gamma ^2 \cA  \left(2 q^2 Z_X-\left(2 X Z_X+Z\right) \left(2 \cA-r \cA' Z\right)\right)}{2 r    \Znew}K\cr
    &+&\frac{\gamma ^2 \cA^2 }{Z}K'+\frac{Z^2 \Ynew }{   \Znew}\dot{K} +
    \frac{2 {\Ynew}^2}   {q r    \Znew} \delta \dot\phi +\frac{2 \gamma ^2
\cA^2 \Ynew }{q r Z^3}\delta \phi^{\prime}\,,
\end{eqnarray}
where we have introduced the notation
\begin{eqnarray}
    {\Znew}= Z^3 -Z_X \gamma ^2 \cA \psi '^2 \, ,\qquad {\Ynew}= q \gamma^2  \frac{\cA Z_X \psi'}{Z}.
\end{eqnarray}
In terms of the gauge-invariant combinations \eqref{Horndeski_gauge_invariant1}–\eqref{Horndeski_gauge_invariant3}, this corresponds to the expression
\begin{equation}
\mv=\frac{{\Ynew}^2}{r   \Znew}I_0+\frac{2 \gamma ^2 \cA  \Ynew }{r Z^3}I_1+\frac{\gamma ^4 \cA^2   \Znew}{r Z^6}I_2 \, .
\end{equation}

As we have explained in Section \ref{perturbations section}, the master variable enters in an equation of the form \eqref{EqE1} whose  coefficients can be determined explicitly although their expressions are too cumbersome to present here.

The next step consists in finding the system of equations $\{\Ne,\alg\}$  \eqref{EqE1}-\eqref{EqE2}. Following the strategy described in Section \ref{perturbations section}, we find that
\begin{equation}
 \Ne=  \mathscr{E}_{tt}
   +\frac{\cA \Ynew }{Z \left(\cA \left(2 X Z_X+Z\right)-q^2 Z_X\right)}\mathscr{E}_{tr}\, ,
\end{equation}
\begin{equation}
 \alg=  \mathscr{E}_{tr}
   +\frac{ \cA \Ynew}{Z \left(\cA \left(2 X Z_X+Z\right)-q^2 Z_X\right)}\mathscr{E}_{rr}\, ,
\end{equation}
where we have replaced $H_2$ by its expression in terms of $\sigma$ using \eqref{mv_Horndeski}.

The two equations \eqref{EqE1} and \eqref{EqE2} enable us to find a second order equation \eqref{def_second_order_equation} for the master variable $\sigma$  provided that  the determinant \eqref{determinant_def} does not vanish. The determinant can be written as
\begin{equation}
\label{D_Horndeski}
\begin{aligned}
    \Det=&-\frac{\gamma ^2 q^4 \cA^3 \left[F \left(2 X Z_X+Z\right)+Z^2\right] \left(q^2-2 X \cA\right) }{8 r^5 X^4 Z^4 X' \left[Z \cA-Z_X \left(q^2-2 X \cA\right)\right]{}^4} \, \hat\Delta\,,
\end{aligned}
\end{equation}
where $\hat\Delta$ is given by
\begin{eqnarray}
\hat\Delta & =& 
     F \left(2 X Z_X+Z\right){}^2 \left[\gamma ^2 q^2 Z \left(r X'+2 X\right)+2 \gamma ^2 q^2 r X Z_X X'-4 X^2 Z^3\right] \nonumber \\
    &+&Z^2 \left[4 \gamma ^2 q^2 X Z \left(Z_X \left(r X'+X\right)-r X X' Z_{XX}\right)+Z^2 \left(\gamma ^2 q^2 \left(r X'+2 X\right)-24 r X^3Z_X^2 X'\right)\right.\nonumber\\
    &+ &\left. 16 \gamma ^2 q^2 r X^2Z_X^2 X'-8 X^3 Z^3 \left(Z_X-r X' Z_{XX}\right)-4 X^2 Z^4\right] \, .
\end{eqnarray}
This expression shows that the determinant does not vanish in general. Hence, $\sigma$ satisfies a second order equation of the form \eqref{def_second_order_equation}.

Finally, to transform the second-order equation into a Schrödinger-like equation, we redefine the time coordinate according to \eqref{def_tstar} where $W$ is explicitly given by\footnote{ As in the previous Section where $W$ simply reduces to \eqref{timechange_p}, here the function $W$ depends also on the choice of the branch we make for $\psi$. We choose the branch such that $t_*$ is regular at the horizon.}
\begin{equation}
\label{W_Horn}
    W = q\frac{ 2 X Z^3-\gamma ^2 Z \left(r \cA X'+q^2\right)-2 \gamma ^2 r X \cA Z_X X'}{\gamma  \cA^2 \psi' \left(\gamma ^2 q^2-2 X Z^2\right)} \, .
\end{equation}
At this stage, we have all the ingredients to compute the Schrödinger-like equation for the monopole with the explicit Schrödinger radial coordinate $r_*$ and the effective potential $\Veff$ \eqref{Veffgeneral}. The general expressions, however, are too lengthy to be included in this paper. We therefore provide a GitHub page where the full expressions for all these quantities can be downloaded in their most general form.

\section{Generic conformal-disformal  transformations}
\label{sec.ConfDis}
In this Section, we review some properties of conformal-disformal transformations that are needed in the article. 
\subsection{Transformations in the space of theories}
As shown in \cite{BenAchour:2016cay}, a conformal-disformal  transformation  of the  metric,
\begin{align}
\label{disformal_transf}
    \tilde g_{\mu\nu}= \cCo(X,\phi) \, g_{\mu\nu} + \cDis(X,\phi) \, \partial_\mu \phi\, \partial_\nu \phi\,,
\end{align}
where $C$ and $D$ are functions of $X$ and $\phi$,
induces an internal map in the space of DHOST theories. 
Indeed, any DHOST theory governed by an action $\tilde S[\tilde{g}_{\mu\nu},\phi]$ is related to another DHOST theory governed by the action
\begin{eqnarray}
\label{disactions}
    S[g_{\alpha\beta},\phi] \equiv \tilde{S}[\tilde{g}_{\mu\nu}(g_{\alpha\beta},\phi),\phi]\,.
\end{eqnarray}
The explicit disformal transformations between quadratic DHOST theories can be found\footnote{Beware that $X$ in those references is defined as $X_{\rm other}\equiv \partial_\mu \phi \, \partial^\mu \phi$.} in the appendix D of \cite{Langlois:2020xbc} (correcting some typos in \cite{BenAchour:2016cay}). 
Here, we  just recall the expressions of the new functions $\tilde \Ftwo$ and $\tilde A_1$, given respectively by
\begin{eqnarray}
\label{disformed_F2}
\tilde \Ftwo & = & 
\frac{\Ftwo}{\cCo (1 -2X \cDis/\cCo)^{1/2}}
 \, ,\\
 \label{disformed_A1}
\tilde A_1 & = & \left( 1 -2X {\cDis}/{\cCo} \right)^{3/2} \left(A_1 - \frac{\cDis}{\cCo-2 X\cDis } \Ftwo\right) 
\,,
\end{eqnarray}
while the relation between $P$ and $\tilde{P}$ is the same as in \eqref{disformed_F2}.
It is also useful to recall the relation between $X$ and $\tilde{X}$,
\begin{align}
\label{disformed_X}
    \tilde{X}=\frac{X}{\cCo-2X\cDis}\,,
\end{align}
which can be inverted (not always explicitly) and substituted into \eqref{disformed_F2} and \eqref{disformed_A1} so that $\tilde \Ftwo$ and $\tilde A_1$ are expressed as functions of $\tilde{X}$. 

\medskip

Let us conclude this subsection with three short remarks. 
\begin{itemize}
\item First note that
 the transformations \eqref{disformed_F2} and \eqref{disformed_A1} are well defined only if $C$ and $D$ satisfy the condition 
\begin{eqnarray}
\label{CondDis}
    1-2X D/C \, > \, 0 \, , 
\end{eqnarray}
which in turn ensures that $\tilde{X}$ has the same sign as $X$ when, in addition, $C>0$. 
\item Any DHOST theory of type Ia can be related, via disformal transformations, to a Horndeski theory, with the well-known interesting property that all the equations of motion are second order partial differential equations. 
\item We stress that, while the seed and image theories are equivalent theories, provided the disformal transformation is invertible\footnote{The theories are inequivalent if the disformal transformation is non-invertible, which leads to mimetic-like theories, as discussed in \cite{Langlois:2018jdg} for instance.}, they become physically inequivalent  as soon as one introduces matter and assumes it is  minimally coupled to their respective metric. In particular, in contrast with purely conformal transformations, disformal transformations (with a non trivial $D$ function)  lead to a new metric with a different light cone structure. 
\end{itemize}

\subsection{Transformation of the metric}

Let us now  compute explicitly the disformed metric $\tilde{g}_{\mu\nu}$ \eqref{disformal_transf} obtained from the spherically symmetric metric  \eqref{line_element} and  the scalar field \eqref{phiback}. We  obtain  the new static and spherically symmetric metric 
\begin{eqnarray}
\mathrm{d}{\tilde s}^2 = -(\cCo \cA - \cDis q^2)  \left( \mathrm{d}{t} - \cDis \frac{ q \psi'}{\cCo \cA - \cDis q^2} \mathrm{d}{r} \right)^2 + \cCo \left( \frac{1}{\cB} +  \frac{\cDis \cA\,  \psi'^2}{\cCo \cA - \cDis q^2}\right)  \mathrm{d}{r}^2 + 
\cCo\,  r^2  \mathrm{d}{\Omega}^2 \, .
\end{eqnarray}
We can easily diagonalise the above metric by introducing the new
time variable
\begin{equation}
\label{t_disf}
    \tilde{t}=t-q\int\frac{\cDis\psi' }{\cCo \cA - \cDis q^2}\mathrm{d} r\,,
\end{equation}
which leads to the disformed line element
\begin{eqnarray}
d\tilde{s}^2 = -\tilde\cA(r) \, \mathrm{d}{\tilde{t}}{}^2+\frac{\mathrm{d} r^2}{\tilde\cB(r)} +\tilde\cC(r) \, \mathrm{d} \Omega^2 \,  ,
\end{eqnarray}
with the new metric coefficients
\begin{eqnarray}
 &&\tilde\cA= \cCo \left(\cA - q^2{\cDis}/{\cCo }\right) \, , \quad \frac{1}{\tilde \cB} =  \cCo \, \frac{\cA}{\cB} \left(   \frac{1 -2X{\cDis}/{\cCo }}{\cA - q^2{\cDis}/{\cCo }}\right) \, , \quad
\tilde\cC= \cCo r^2 \, .
\label{disformed_metric}
\end{eqnarray}
In terms of the new time coordinate \eqref{t_disf}, the scalar field is expressed as  
\begin{equation}
\label{phi_dis}
    \phi =q \tilde{t}+\int\mathrm{d} r  \; \frac{\cA \,  \psi'}{\cA -q^2{\cDis}/{\cCo } }\,.
\end{equation}
By construction, the new  metric is a solution of the DHOST theory $\tilde{S}$ obtained from the original one $S$ by a disformal transformation  \eqref{disactions}. 

Furthermore, as can be seen from the expressions of the metric coefficients \eqref{disformed_metric}, the condition \eqref{CondDis} together with $C>0$, also ensures that $g_{\mu\nu}$ and $\tilde{g}_{\mu\nu}$ keep the same global signature, even if  the signs of $\cA$ and $\cA - q^2{\cDis}/{\cCo }$ might differ (as will be sometimes the case in our examples). 

\subsection{Disformal transformation of $p=2$ theory into a Horndeski theory}
\label{Detailsofp2toHorn}
In this subsection, we give some details of the disformal transformation \eqref{D_H_p2} that maps the theory \eqref{P_n} with $p=2$ into a Horndeski theory.  

We start by computing the
expression of $\XH$ in terms of $X$ 
\begin{align}
\XH &= \frac{X}{1-3\alpha X^2}
\;\;\Longrightarrow\;\;
\left\{
\begin{aligned}
X &= \frac{ {\Upsilon}-1}{6 \alpha \XH} \quad \text{if} \quad \alpha >0 \\
 X &=-\frac{{\Upsilon}+1}{6 \alpha \XH}\quad \text{if} \quad \alpha <0 
\end{aligned}
\right\} 
\quad \text{with} \;\; \Upsilon = \sqrt{1 + 12 \alpha \tilde{X}^2} \,.
\end{align}
It is easy to check that $\Upsilon$ is a well-defined real-valued function since
\begin{eqnarray}
  1 + 12 \alpha \tilde{X}^2  = \left(\frac{1+3\alpha X^2}{1-3\alpha X^2} \right)^2\, .
\end{eqnarray}
Then, the relations \eqref{disformed_F2} and \eqref{disformed_A1} allow us to determine the functions appearing in the Horndeski action. For simplicity, we assume $\alpha>0$ (the case $\alpha<0$ is presented in Appendix B of \cite{Charmousis:2025xug}) and we obtain  \eqref{def_N0H}.

\section{More on the positivity of the Schrödinger operator}
\label{sec.GeneralBoundaryConditions}
In the main text we have chosen for the function $S$ the particular solution \eqref{Sfunction_Hairy}, verifying \eqref{V_S} for $\nth>2$ .
One can consider the more general function
\begin{align}
S=\frac{r^2+1}{2(2p-1)\, r\, n} &\left[\frac{\xi_{\nth}  \left(r^2+1\right)^{1-p} \left(4r^3+3 r( 1-\cA)+2 \mu -2 c_S \xi_{\nth} \right)}{r (\cA-1)+2 \mu -2 c_S \xi_{\nth} } \right. \nonumber \\
&-\left.\frac{\left((5-4 p) r^2+1\right) \cA+\left(r^2+1\right) \left(4 (p-2) r^2-1\right)}{2 r^2}\right] \,,
\end{align}
which depends on the constant of integration $c_S$. This function   is well defined for $r\in \, ]0,+\infty[$, and therefore for $0<r_*<\rmax$ whenever $c_S\in \mathbb{R}\setminus\left]-\frac{ \sqrt{\pi } \Gamma \left(p-\frac{3}{2}\right)}{4 \Gamma (p)},0\right[$.
None of the asymptotic behaviors described between \eqref{psiandS_asymptot} and \eqref{psi_asympt} are affected by this new choice of integration constant
while the asymptotic expansion of the function $S$ is now given by
\begin{eqnarray}
\label{S_asympt1}
    S(r)= \sqrt{\frac{\mu }{4p-2} } r^{-5/2}\left(\frac{1}{2}+\left(\frac{5}{2}-2 p\right) r^2+r^3\left(\frac{7 (2 \xi_{\nth} +3)}{24 \mu }-\frac{2 }{c_S}\right)+ {\cal O}(r^4)\right) ,
\end{eqnarray}
Using now \eqref{n_asympt}, when $r \rightarrow 0$, we obtain 
\begin{eqnarray}
     \overline{\Psi} \Dope\Psi \, = -\; \sqrt{\frac{2\mu}{2p-1} } \left[ \vert {c}_-\vert^2  \left(\frac{1}{c_S}-\frac{7  (2 \xi_{\nth} +3)}{8 \mu }\right) -3\,  {{c}_+} \overline{c}_-\right] + {\cal O} (r) \, .
\end{eqnarray}
We impose the following boundary condition at $r = 0$:
\begin{eqnarray}
\label{boundcondalphaC}
   c_+ \, \cos \theta \,  + \,  c_- \, \sin\theta \,  = \; 0 \, ,
\end{eqnarray}
where $\theta$ is an arbitrary angle that parametrizes the choice of the self-adjoint extension. The contribution of the boundary term at the origin can be expressed in terms of the asymptotic coefficients as,
\begin{equation}
    {\Bterm}_0 =-\sqrt{\frac{2\mu}{2p-1}}|c_-|^2 \left[    \frac{1}{c_S}-\frac{7(2\xi_{\nth}+3)}{8\mu}  +3\tan\theta   \right] .
\end{equation}
Therefore, we find that the boundary contribution is positive provided that $\theta$ satisfies
\begin{equation}
\tan \theta<-\frac{1}{3}\left(\frac{1}{c_S}-\frac{7(2 \xi_{\nth}+3)}{8 \mu}\right)\, .
\end{equation}
This selects a generic subset of self-adjoint extensions for which the Schrödinger operator is positive.

\end{appendices}
\newpage
\bibliographystyle{utphys}
\bibliography{References}

\end{document}